\documentclass[prl,twocolumn,floatfix,reprint,nofootinbib]{revtex4}
\usepackage{graphicx,amssymb,amsmath}
\usepackage{stmaryrd}
\usepackage{hyperref}

\begin{document}

\title{Transitions in spatial networks}

\author{Marc Barthelemy}
\email{marc.barthelemy@ipht.fr}
\affiliation{Institut de Physique Th\'{e}orique, CEA, CNRS-URA 2306, F-91191, 
Gif-sur-Yvette, France}
\affiliation{CAMS (CNRS/EHESS) 190-198, avenue de France, 75244 Paris Cedex 13, France}

\begin{abstract}

  Networks embedded in space can display all sorts of transitions when
  their structure is modified. The nature of these transitions (and in some cases
  crossovers) can differ from the usual appearance of a giant
  component as observed for the Erdos-Renyi graph, and spatial
  networks display a large variety of behaviors. We will discuss here
  some (mostly recent) results about topological transitions,
  `localization' transitions seen in the shortest paths pattern, and
  also about the effect of congestion and fluctuations on the structure of optimal
  networks. The importance of spatial networks in real-world
  applications makes these transitions very relevant and this review
  is meant as a step towards a deeper understanding of the effect of
  space on network structures.
\end{abstract}


\maketitle

\section{Introduction}

Even if networks became very fashionable these last two decades, there
is even a longer history between networks and transitions. In their
seminal paper Erdos and Renyi \cite{Erdos:1960} showed at
the end of the $50$'s the existence of a transition in random graph
when the number of links increases. This Erdos-Renyi (ER) model
\cite{Erdos:1959,GIlbert:1959} exists
independently from space, but very early, Gilbert introduced a simple
model of what we now call a spatial network \cite{Gilbert:1961}. In its
simplest version, nodes in this model lie in the 2-dimensional space
and are connected if their euclidean distance is less than a given
threshold. This model also displays a transition for a specific value
of the average degree and, similarly to the Erdos-Renyi model, this
transition is of the percolation type characterized by the sudden appearance of a
giant component connecting most nodes. Both these models triggered a
lot of attention and many results were obtained, mosty by
mathematicians (see for example \cite{bollobas:1985,Penrose:2003}). More recently,
networks embedded in space -- spatial networks  -- became of interest as they describe many important
systems in the real-world, ranging from transportation networks,
infrastructures such as power grids, to biological structures such as
the brain or veination patterns in leaves. For most of these
networks, nodes are located in a two-dimensional space and links
represent physical connections (cables, wires, axons, etc.). The
central effect of space is then to associate a cost to the length
of links and to the degree of a node: a long link is costly and is
usually compensated by some advantage, and spatial constraints limit
the degree and prohibits the appearance of broad distributions. It is this interplay that leads to
the richness of these structures. 

New network models were proposed that integrate the effect of space
and are based in general on different mechanisms, from variants of the
preferential attachment, to greedy models or optimal problems. Many of
these models display various types of transitions or crossovers that
we will describe here, but a clarification is needed at this point. In
statistical physics, transitions are usually well defined: they
describe how the system goes from one equilibrium phase to another
one. These phases are usually described by different symmetries and
the transition is characterized by a non-analytic behavior in the
thermodynamic limit. We can also observe crossovers which in general
involve a typical scale and which separate different regimes. The
change is then more progressive and we don't have properties such as
the loss of symmetry or non-analyticity. In `non thermodynamical'
(and in general `complex') systems, the situation is not always clear and the meaning
of the word transition is less strict. The absence of a free energy
function for these sytems leaves the freedom to characterize its
behavior by many other quantities that can display a change more or less
abrupt. By accepting that the word transition can be applied to these
cases, it can encompass many processes where a particular quantity
undergoes a change when varying a control parameter. The change can be
abrupt and would then correspond to the usual idea of transition with
the existence of a critical value for the control parameter. In some
cases however, the system can interpolate continuously between two
extremes and it is probably better to speak of a crossover (that could
take place in the infinite size limit and does not describe a change
when the size is increasing), and is described in general by a
scaling function. In this review we will mostly speak about abrupt
changes between different regimes but we will also mention some
examples of this sort of crossover.

Understanding these transitions (and crossovers) will help us to
identify the control parameters for these systems and eventually to
improve the modelling of real-world spatial networks. This review is
thus an attempt to bring together these different models under the
unifying concept of transition which might help us to construct a
consistent framework for spatial network modelling.

We will review some classes of models recently developped and by no
means we intended to be exhaustive, but rather tried to focus on
results that might trigger further studies, or provoke thoughts and
modelling of real-world networks. Also, we will mostly focus here on
transitions about the structure of graphs and not transitions for
processes that take place on networks. We will start this review
with the percolation type of transition. This phenomenon is by now
well-known and we won't insist on this part as many books and reviews already
exist (see for the example \cite{Grimmett:1999} and for a more
statistical physics oriented book see \cite{Stauffer:2014}). We will
basically discuss the random geometric graph for which many results
exist \cite{Penrose:2003} and which represent a good example of a simple and rich model of
a spatial network.

We will then discuss `topological transitions' in various models of
graphs that are characterized by a change in the structure of the
graph, as measured by a specific quantity such as the
average shortest path that can display either a small- or large world
behavior. We illustrate this effect on the Watts-Strogatz model and
its d-dimensional variant, and also on a spatial variant of the
preferential attachment where sharp transitions were observed. We will
also discuss this sort of transition for a class of greedy models
based on a cost-benefit analysis and that represent a good candidate
for modelling various real-world systems.

In the next part we will discuss some sort of `localization'
transition where modifications are seen in the spatial organization of
shortest paths, in particular with the concentration of bottlenecks
(i.e. nodes with a large centrality) in a small region of space. In
particular, for a toy model consisting of a ring and branches we
observe a transition for which the ring becomes central in the
organization of shortest paths. We will also discuss a more general
model which include randomness and that exhibits this localization
type of transition when the density of links increases.

Finally we will discuss transitions in optimal networks, a very
important and well-developped topic. After a short discussion about
transitions between classical optimal trees, we will discuss some new
results about the effect of congestion on topological transitions in optimal trees. We
will end this part with a discussion about the effect of noise and
fluctuations in the formation of loops.

\section{Percolation-type transition}

For this type of transition, a giant cluster that spans the whole
system appears suddenly when we increase the fraction of existing
links. In the standard percolation problem on a lattice (see
\cite{Stauffer:2014}) the control parameter is the probability of
presence of a link and there is a sharp threshold that depends on the
underlying lattice. For graphs, the first instance of this type of
transition was described in the celebrated paper by Erdos and Renyi
\cite{Erdos:1960}. In this class of random graph models, we typically
have a set of $N$ vertices and a probability $p$ that any pair
is connected. For this model the degree distribution is binomial
\begin{align}
P(k)=\binom{N}{k}p^k(1-p)^{N-k}
\end{align}
which in the case of large networks with $\langle k\rangle =Np=\mathrm{const.}$
for $N\to\infty$, converges to the Poisson distribution
\begin{align}
P(k)=\frac{\langle k\rangle^k}{k!}\mathrm{e}^{-\langle k\rangle}
\end{align}
where $\langle k\rangle$ is the average degree of the graph. The now
standard result obtained by Erdos and Renyi in their 1960 paper
\cite{Erdos:1960} is the evolution of this graph when $p$ is
varied. For large $N$ the only parameter is the average degree
$\langle k\rangle$ and they obtained a transition for
$\langle k\rangle_c =1$. More precisely, they could show that:
\begin{itemize}
\item{} For $\langle k\rangle<1$, clusters have a typical size of
  order ${\cal O}(\log N)$.
\item{} For $\langle k\rangle =1$, there is a giant cluster of size
  scaling as $N^{2/3}$. 
\item{} For $\langle k\rangle>1$, there is a giant component of
  extensive size (a fraction of $N$). In addition, no other component
  is larger than ${\cal O}(\log N)$.
\end{itemize}

Other results about the connectivity of this graph can be
obtained. For example, if $p<\log N/N$, the graph will contain
isolated vertices and above this sharp threshold the graph is almost
surely connected. Many other results were subsequently obtained for
this graph and we refer the interested reader to the book
\cite{bollobas:1985}.


In the case of spatial networks, one of the first example that
displays this type of percolation-type transition is the random
geometric graph \cite{Gilbert:1961}, (also called the unit disk graph)
which became an important model with many applications. The random
geometric graph is obtained from a random distribution of points in
the plane and a geometric rule for connecting these points and
creating edges. The simplest case is when a proximity rule is used
which states that nodes only within a certain distance are
connected. There is an extensive mathematical litterature (see the
book \cite{Penrose:2003} and references therein) on these graphs and
they were also studied by physicists in the context of continuum
percolation (see for example
\cite{Balberg:1985,Quantanilla:2000}). This process extends usual
percolation theory to continous space where shapes are randomly
positioned and can overlap \cite{Gawlinski:1981,Meester:1996}.

Random geometric graphs are probably the simplest models of spatial
networks and they can be used to model or understand many real-world
situations. This is the case of wireless networks, smart-grids,
disaster relief, etc. (see \cite{Dettmann:2016} and references
therein). In particular, in ad-hoc networks~\cite{Nemeth:2003} where
users communicate by means of short range radio devices that can
communicate with each other if their distance is less than their
transmission range. The set of connected devices can be used to
propagate information over a longer distance by going from the source
to the destination hoping through intermediate nodes. If there is a
large density of nodes, alternate routes are even available which
allows to split the information into separate flows. Usually, the
users are mobile and the network evolves in time and it is important
to understand the condition for the existence of a giant cluster. The
percolation threshold and other quantities have then a direct interest
in this type of applications.

\begin{figure}[h!]
\centering
\includegraphics[width=0.5\textwidth]{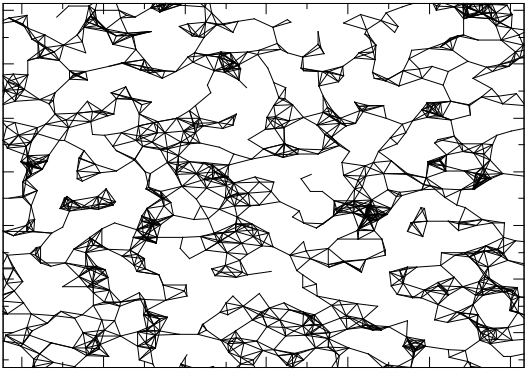}
\caption{ Example of a 2d random geometric graph obtained with a
  density and radius such that the average degree is $\langle
  k\rangle\approx 6$. Note that this is an example of a non-planar
  spatial network.}
\label{fig:dall}
\end{figure}

This model was introduced by Gilbert \cite{Gilbert:1961} who assumes
that $N$ points are randomly located in the plane and have each a
communication range $R$. This also could be seen as a system of disks
(or spheres in dimension $d$) of radius $r$. Two nodes are connected
by an edge if they are separated by a distance less than $R$ (or $2r$
for the distance between the centers of the disks). We show an example
of such a network in Fig.~\ref{fig:dall}.

If we denote by $\rho=N/A$ the density of nodes in area $A$ in the
$d=2$ case, the average degree is given by
\begin{align}
\langle k\rangle=\rho\pi R^2
\end{align}
Similarly to the Erdos-Renyi random graph, there are different
quantities that we can compute. In particular, there is a percolation
transition for a critical density and another transition to full
connectivity (ie. there are no isolated nodes).

Most studies are conducted in the limit $N\to\infty$ which can be
achieved in different ways. A first way is to consider a finite total
area $A$, an increasing density $\rho$ but a range $R$ that decreases with
$N$ such that $\langle k\rangle$ is fixed. Another way -- mostly
considered by mathematicians (see for example \cite{Penrose:2003}) is to study this
limit by considering a fixed $R$ and a fixed density but an area that
varies as $N/\rho$.

It has been demonstrated (see the book \cite{Penrose:2003} and references
therein) that for large $N$ there is a critical density (at fixed $R$
and area given by $N/\rho$) below which we have small
components of typical size ${\cal O}(1)$ and of largest size $\sim\log N$, and a giant cluster
of size $\sim N$. Gilbert \cite{Gilbert:1961} discussed already the probability to
belong to an infinite cluster $P_\infty$ and found a critical average
degree 
\begin{align}
\langle k\rangle_c\approx 4.7
\end{align}
and later the authors of \cite{Dall:2002} found $\langle k\rangle_c\approx 4.52$ (for
$N\approx 10^6$ nodes). Using analytical techniques the authors of
\cite{Balister:2005} could show that with $99.99\%$ confidence 
\begin{align}
4.508 \leq \langle k\rangle_c \leq 4.515,
\end{align}
which is consistent with the numerical bounds found in \cite{Quantanilla:2000}. 
\begin{align}
4.51218 \leq\langle k\rangle_c\leq 4.51228
\end{align}
We can note here that this value is much larger than its counterpart
obtained for the ER graph. It seems here that spatial constraints
impose a larger average degree in order to create a giant cluster.

\section{Topological transitions}

For spatial networks, a fundamental quantity is the distribution
$P(\ell_1)$ of the length $\ell_1$ of links. When this distribution is peaked with a
fast decaying tail, we can expect a lattice-like behavior for most
measures on this network: the average shortest path will be `large'
(in general scaling as a power of the number $N$ of nodes) and most
quantities such as the clustering are much larger than their random
counterparts (ie. computed for a graph without space such as the ER
graph). If in contrast, the distribution $P(\ell_1)$ is broad, we have a
few long links that can be of the order the system size and the most
relevant quantities will be affected: the average shortest path will
be much smaller than the lattice-like case (typically varying as
$\log N$). For some models, varying a parameter can have a dramatic
effect on $P(\ell_1)$ and we can then observe a transition between
different classes of networks with different large-scale behavior. We
will speak here of a `topological transition' and we will present two
examples of such models that are based on preferential attachment and
on a cost-benefit analysis.

\subsection{Watt-strogatz small-worlds}

\begin{figure}[h!]
\centering
\includegraphics[angle=0,width=0.4\textwidth]{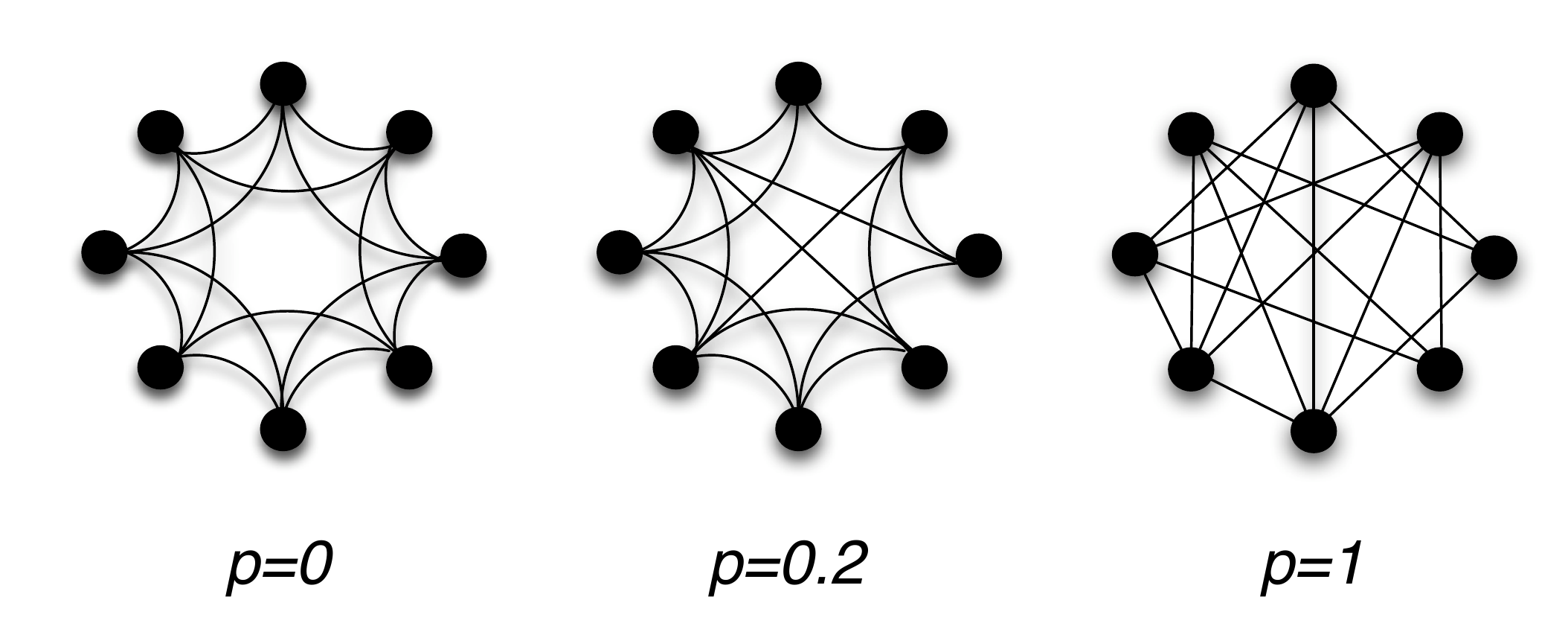}
\caption{ Construction of the Watts-Strogatz model for $N=8$ nodes. At
  $p=0$ each node is connected to its four nearest neighbors and by
  increasing $p$ an increasing number of edges is rewired. Adapted
  from Watts and Strogatz \cite{Watts:1998}. }
\label{fig:sw}
\end{figure}

Already in $1977$, spatial aspects of the small-world problem were
considered by geographers in \cite{Stoneham:1977} but we had to wait
until $1998$ when Watts and Strogatz (WS) proposed a simple and
powerful network model \cite{Watts:1998} which incorporates both a
spatial component and long-range links. This model is obtained by
starting from a $d=1$ regular lattice and by rewiring links at random
with a probability $p$ (see Fig.~\ref{fig:sw}).

The degree distribution of this network has essentially the same
features as the ER random graph, but the clustering coefficient and the
average shortest path depend crucially on the amount of randomness
$p$. The average clustering coefficient has been shown to behave as
\cite{Barrat:2000}
\begin{align}
\langle C(p)\rangle\simeq 
\frac{3}{4}
\frac{\langle k\rangle-2}{\langle k\rangle -1}
(1-p)^3
\end{align}
The average shortest path has been shown to scale as \cite{Barthelemy:1999,Erratum_Barthelemy:1999}
\begin{align}
\langle \ell\rangle \sim N^*{\cal F}\left(\frac{N}{N^*}\right)
\end{align}
where the scaling function behaves as 
\begin{align}
{\cal F}(x)\sim
\begin{cases}
x\;\;\;\;\;&{\rm for}\;\; x\ll 1\\
\ln x\;\;&{\rm for}\;\;x\gg 1
\end{cases}
\end{align}
We thus observe a crossover between different network phases when the
number of nodes increases. This change from a large to a small-world is typically a crossover
between two structures and strictly speaking in the large $N$ limit we
are always in the small-world regime. 

These results show that the WS network can be seen as clusters of typical size $N^*$ connected by
shortcuts. The crossover size scales as $N^*\sim 1/p$
\cite{Barthelemy:1999,Erratum_Barthelemy:1999,Barrat:2000} which
basically means that the crossover from a large-world to a small-world
occurs for an average number of shortcuts of the order of one
\begin{align}
N^*p\sim 1
\label{eq:sw}
\end{align}

We note that historically, the interest of these networks is that they can simultaneously present
some features typical of random graphs (with a small-world behavior
$\langle\ell\rangle\sim \log N$) and of clustered lattices with a large average
clustering coefficient (while for the ER random graph we have $\langle C\rangle
\sim 1/N\ll 1$).

\subsection{d-dimensional generalization}

One of the first variants of the Watt-Strogatz model was proposed in
\cite{Jespersen:2000,Kleinberg:2000,Sen:2001} and was subsequently
generalized to higher dimensions $d$ \cite{Sen:2002b}.
\begin{figure}[h!]
\centering
\includegraphics[width=0.5\textwidth]{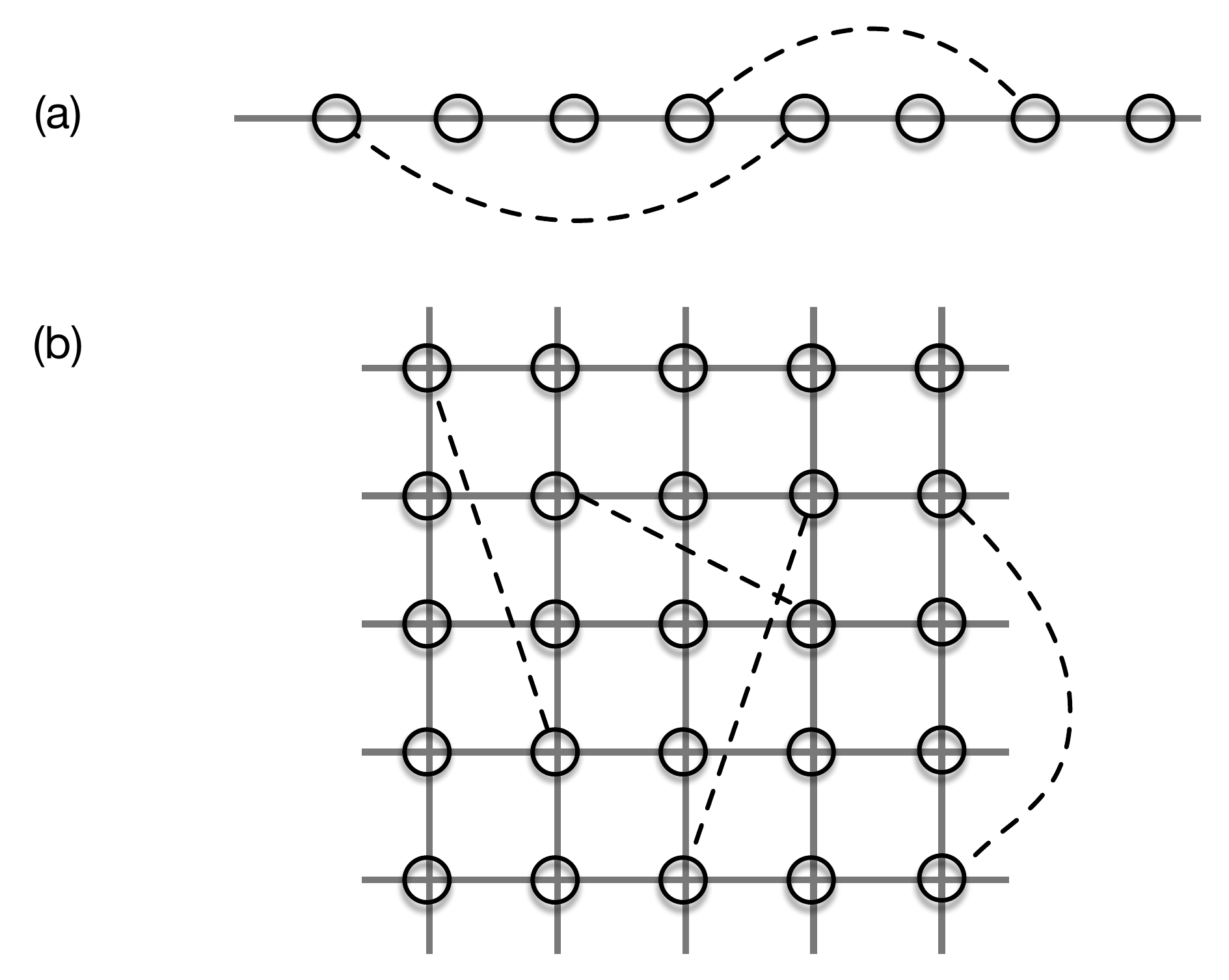}
\caption{ Schematic representation of a spatial small-world in (a) one
  dimension and (b) two dimensions. The dashed lines represent the
  long-range links occurring with probability $q(\ell)\sim
  \ell^{-\alpha}$. Figure inspired from \cite{Sen:2002b}.}
\label{fig:sswnet}
\end{figure}
In this variant (see Fig.~\ref{fig:sswnet}), nodes are located on a
regular lattice in $d$-dimensions (with periodic boundaries or not). 
For each node, we add a shortcut with probability $p$ which implies
that on average there will be $pN$ additional shortcuts. We now define
two differents models A and B according to the choice of the length
distribution of shortcuts. For the model A, the length $r$ of
these links follows the distribution
\begin{align}
q_A(r)\sim r^{-\alpha}
\end{align}
where $\alpha$ is independent from the dimension $d$.
The main idea for justifying this choice is that if shortcuts have to
be physically realized there is a cost associated with their length
and thefore a probability that decreases with the length. In this
model A the effect of dimensionality is limited: we choose a link
length distributed according to $q_A(r)$ and this is independent
from the dimension $d$ of the underlying lattice. In contrast, in
model B (considered for example by Kleinberg \cite{Kleinberg:2000}) we
first choose a node $x$ in the d-dimensional lattice and an endpoint
$y$ with probability proportional to $|x-y|^{-\alpha}$. The length
distribution is then
\begin{align}
q_B(r)\sim r^{-\alpha}r^{d-1}
\label{eq:qb}
\end{align}
where the last term comes from the volume element
$r^{d-1}\mathrm{d}r$. The conceptual difference between models A and B is that in A we
choose a distance and then a vector with this length. In the model B
we are choosing directly a vector with a endpoint density $q_B$. We
will therefore discuss these models separately but if we substitute
$\alpha\to\alpha-d+1$ in model A we should recover the results
of model B \footnote{I thank P. Grassberger for discussions on this
  point}.  There is then a correspondance between the regimes in the two 
models and we give it in Table~\ref{table0}.
\begin{table}[h!]
\begin{center}
 \begin{tabular}{|c |c|} 
 \hline
Model A & Model B \\ 
[0.5ex] 
 \hline\hline
$\alpha<1$  & $\alpha<d$ \\ 
 \hline
$1<\alpha<d+1$ & $d<\alpha<2d$ \\
 \hline
$d+1<\alpha$ & $2d<\alpha$ \\ [1ex] 
 \hline
\end{tabular}
\end{center}
\caption{Correspondance between the regimes for models A and B
  obtained by replacing $\alpha$ with $\alpha-d+1$.}
\label{table0}
\end{table}

\subsubsection{Model A}

It is clear that if $\alpha$ is large enough, the shortcuts will be
small and the behavior of the average shortest path
$\langle\ell\rangle$ will be `spatial' with
$\langle\ell\rangle\sim N^{1/d}$. On the other hand, if $\alpha$ is
small enough we can expect a small-world behavior characterized by
$\langle\ell\rangle \sim \log N$.  Various studies
\cite{Jespersen:2000,Moukarzel:2002,Sen:2002b} discussed the existence
of a threshold $\alpha_c$ separating the two regimes, small- and
large-world and we will also discuss this result in the next section
about preferential attachment and space. Here, we will follow the
discussion of \cite{Petermann:2005,Petermann:2006} who studied
carefully the behavior of the average shortest path.  The probability
that a shortcut is `long' is given by
\begin{align}
P_c(L)=
\frac{\int_{(1-c)L}^{L}q_A(r)\mathrm{d}r}
{\int_{\ell_m}^Lq_A(r)\mathrm{d}r}
\end{align}
where $c$ is small but non-zero and where the lower cut-off $\ell_m$
goes in general to zero with $N$ (typically, $\ell_m\sim
1/\sqrt{\rho}\to 0$
where $\rho$ is the density increasing with $N$). Performing these integrals we obtain
\begin{align}
P_c(L)=
\frac{L^{1-\alpha}-(1-c)^{1-\alpha}L^{1-\alpha}}
{L^{1-\alpha}-\ell_m^{1-\alpha}}
\end{align}
We denote by $p^*$ the critical fraction of shortcuts and according to
the discussion above about the condition for having a small-world
(Eq. \eqref{eq:sw}), the fraction of `long' shortcuts needs to be
large enough and this reads here (with $N\sim L^d$)
\begin{align}
P_c(L)p^*L^d\sim 1
\end{align}
This condition means that if we have a fraction
$p>p^*$ of long shortcuts, the system will behave as a small-world. If
$\alpha<1$, the fraction $P_c(L)\to 1-(1-c)^{1-\alpha}$ when
$\ell_m\to 0$, and in this case, spatial fluctuations do not play a role
and $p^*\sim 1/L^d$. In contrast, if $\alpha>1$, the fraction $P_c(L)$
depends on density fluctuations via $\ell_m$ and we have
$p^*\sim 1/L^{d-\alpha+1}$. To recap, we thus have the following
behavior
\begin{align}
p^*(L)\sim
\begin{cases}
L^{-d}\;\;\;\;\;&{\rm if}\;\; \alpha<1\\
L^{\alpha-d-1}\;\;&{\rm if}\;\;\alpha>1
\end{cases}
\end{align}
(and a logarithmic behavior of the form $\log L/L^d$ for $\alpha=1$). For a given
value of $p$ we thus have a length scale 
\begin{align}
L^*(p)\sim
\begin{cases}
p^{-1/d}\;\;\;\;\;&{\rm if}\;\; \alpha<1\\
p^{1/(\alpha-d-1)}\;\;&{\rm if}\;\;\alpha>1
\end{cases}
\end{align}
a result obtained in \cite{Newman:1999} for the special case $\alpha=0$.
We can then write the following scaling form for the average shortest path
\begin{align}
\langle \ell\rangle=L^*{\cal F}_\alpha\left(\frac{L}{L^*}\right)
\end{align}
where the scaling function varies as
\begin{align}
{\cal F}_\alpha(x)\sim
\begin{cases}
x\;\;\;\;\;&{\rm if}\;\;x\ll 1\\
\ln x\;\;\;\;\;&{\rm if}\;\;x\gg 1
\end{cases}
\end{align}
(it could be a function of the form $(\ln x)^{\sigma(\alpha)}$ with
$\sigma(\alpha)>0$ for $x\gg 1$). For $\alpha<1$, the characteristic
length $L^*\sim p^{-1/d}$ (which is the result for $\alpha=0$) and
implies that there are two regimes (small and large-world). For
$\alpha>1$, the characteristic length behaves as
\begin{align}
L^*(p)\sim p^{1/(\alpha-d-1)}
\end{align}
which displays a threshold value $\alpha_c=d+1$, a value already
obtained in \cite{Sen:2002b}.  In the case $\alpha<\alpha_c$, the
length $L^*(p)$ is finite and there are two regimes depending on the
size $L $versus $L^*$. In contrast when $\alpha\to\alpha_c^-$ this
length diverges and we always have $L\ll L^*$ implying the network is
in the large world regime. In other
words, the links in this case are not long enough and the graph looks
like a lattice at a coarse-grained scale. 

\subsubsection{Model B}

A numerical study of this model was proposed in
\cite{Kosmidis:2008} and numerics for the related model of SIR with long range
infection can be found in \cite{Grassberger:2013a,Grassberger:2013}. We have different 
regimes for this model. First, for $\alpha<d$, the graph is in the small-world regime
with $\langle\ell\rangle \sim \log N$. In the case $\alpha>d$ there are two regimes. First, for
$d<\alpha<2d$ we have a regime of the form
\begin{align}
\langle\ell\rangle\sim (\log N)^{\sigma(\alpha)}
\end{align}
where the exponent has been computed exactly \cite{Biskup:2004}
\begin{align}
\sigma=\frac{\log2}{\log\frac{2d}{\alpha}}\;,
\end{align}
a result which has been confirmed numerically for $d=2$ in
\cite{Grassberger:2013} (and for $d=1$ in \cite{Grassberger:2013a}).
The second regime is obtained for $\alpha>2d$ where the `spatial'
behavior $\langle\ell\rangle \sim N^{1/d}$ is recovered. 

A simple scaling argument was proposed in \cite{Kosmidis:2008} in
order to understand these results and we reproduce it here. It is
based on the average shortcut size given by
\begin{align}
\langle r\rangle=\int_{\ell_m}^L r  q_B(r)\mathrm{d}r
\end{align}
and the largest link size $r_{max}$ defined by the relation
\begin{align}
\int_{r_{max}}^Lq_B(r)\mathrm{d}r\sim \frac{1}{N}
\end{align}
with $q_B(\ell)$ given by Eq.~\eqref{eq:qb}. After a simple calculation one obtains
\begin{align}
\langle r\rangle\sim
\begin{cases}
L\;\;&\alpha<d\\
L^{d+1-\alpha}\;\; &d<\alpha<d+1\\
\mathrm{const.}\;\; &d+1<\alpha
\end{cases}
\end{align}
and
\begin{align}
r_{max}\sim\begin{cases}
L\;\; &0<\alpha<2d\\
L^{\frac{d}{\alpha-d}}\;\;&\alpha>2d
\end{cases}
\end{align}

The picture that emerges from these results is then the following
one. For $\alpha<d$, both the average and the maximum link length is
of order $\langle r\rangle \sim r_{max}\sim L$, and the network is a
small-world characterized by $\langle\ell\rangle\sim\log N$. In the
second regime $d<\alpha<2d$, the average shortcut length is small
($\langle r\rangle\ll L$) but we still have a few very long shortcuts
($r_{max}\sim L$). If we want to connect a pair of points we will have
to do more steps but eventually we will reach a long shortcut that
allows to connect very quickly at destination. The network is
therefore still in a small-world phase with a behavior of the form
$(\log N)^\sigma$. Finally, the third regime ($\alpha>2d$) is
characterized by $\langle r\rangle, r_{max}\ll L$ and the network
is lattice-like with a large-world behavior
$\langle\ell\rangle \sim N^{1/d}$. In addition, it was numerically shown
\cite{Kosmidis:2008} that the clustering coefficient is very small (of
order $1/N$) for $\alpha<d$ and of order unity for $\alpha>d$. The
final picture for this model B is then the following one:
\begin{itemize}
\item{} For $\alpha<d$ the network is a random graph: the clustering
  coefficient is of order $C\sim 1/N$ and the average shortest path
  $\langle\ell\rangle\sim\log N$.
\item{} For $d<\alpha<2d$, the average shortest path is still small ($\langle\ell\rangle\sim (\log N)^\sigma$ with
  $\sigma=1/\log_2(2d/\alpha)$ \cite{Biskup:2004,Grassberger:2013}), and the average clustering
  coefficient $\langle C\rangle$ is large which is a typical feature of small-world
  networks \cite{Watts:1998}.
\item{} For $\alpha>2d$, the average shortest path is of order the
  size of the system $\langle\ell\rangle\sim L\sim N^{1/d}$, a
  behavior typical of a lattice-like network (and $\langle
  C\rangle\sim {\cal O}(1)$).
\end{itemize}

We recap all these different results in the table \ref{table1}.

\begin{table}[h!]
\begin{center}
 \begin{tabular}{|c ||c |c|c |c|} 
 \hline
  & $\alpha<d$ & $d<\alpha <d+1$ & $d+1<\alpha<2d$ & $2d<\alpha$ \\ 
[0.5ex] 
 \hline\hline
$\langle r\rangle$ & $L$ & $L^{d+1-\alpha}$ & const. & const. \\ 
 \hline
 $r_{max}$ & $L$ & $L$ & $L$ & $L^{d/(\alpha-d)}$ \\
 \hline
$\langle\ell\rangle$ & $\log N$ & $(\log N)^\sigma$ & $(\log N)^\sigma$ & $N^{1/d}$ \\
 \hline
 $\langle C\rangle$ & $1/N$ & $\sim{\cal O}(1)$ & $\sim{\cal O}(1)$ & $\sim{\cal O}(1)$ \\ [1ex] 
 \hline
\end{tabular}
\end{center}
\caption{Behavior of various quantities versus $\alpha$ for the model B
  defined by the distribution Eq.~\eqref{eq:qb}: the average shortcut
  length $\langle r\rangle$, the average maximum shortcut length
  $r_{max}$, the average shortest path $\langle\ell\rangle$, and the
  average clustering coefficient $\langle C\rangle$.}
\label{table1}
\end{table}

Finally, we note that model B was also used in
\cite{Kleinberg:2000,Robertson:2006} where the goal was to reach a
given target as fast as possible and using local information only
(i.e. such as the distance to the target for the neighbors). The
average number of hops $\overline{T}$ for reaching a target is then
different from the average shortest path where the whole network is
known. This quantity reaches its minimum for $\alpha=d$ where the
diversity of links is the largest: long links are needed for going
quickly to the neighborhood of the target, and short links are
necessary for reaching the target in the final steps. In general, we
have a scaling of the form $\overline{T}\sim N^s$ where the
exponent depends on both the embedding dimension and $\alpha$
\cite{Kleinberg:2000}:
\begin{align}
s(\alpha)=\begin{cases}
\frac{d-\alpha}{d+1}\;\; &\mathrm{for}\;\;\alpha<d\\
0 &\mathrm{for}\;\;\alpha=d\\
\frac{\alpha-d}{\alpha-d+1}\;\; &\mathrm{for}\;\;\alpha>d
\end{cases}
\end{align}
The exponent $\sigma=0$ corresponds here to a logarithmic behavior of
the form $\overline{T}\sim (\log N)^2$ \cite{Kleinberg:2000}.

\subsection{Preferential attachment and space}

Many networks, including spatial graphs, evolve and grow in
time and understanding the main processes governing this growth and the
resulting structure is crucial in many disciplines ranging
from urban planning to the study of neural networks. There are
essentially three ingredients for growing a network:
\begin{itemize}
\item{} At each time step, one node (or more) is added to the network
\item{} New nodes are located according to a distribution that depends
  in general on the structure of the existing network.
\item{} Once located in space, the new nodes are attached to the
  existing network according to a certain connection rule.
\end{itemize}
We will mostly consider here the case where nodes are uniformly
located, and where the connection rule is governed by a `spatial' variant of
the preferential attachment \cite{Simon:1955,Albert:1999} of the form
\begin{align}
\Pi_{n\to i}\propto k_iF[d(n,i)]
\label{eq:gen}
\end{align}
where $n$ is the new node and $i$ a node that belongs to the existing
network. The function $F$ describes the effect of the euclidean
distance $d(n,i)$ between nodes $n$ and $i$. This is a natural and simple model for spatial
networks as it assumes that when long-range links exist, they usually connect
to hubs -- the well-connected nodes, unless the hub is too far. In order to have
long links, the target node must have a large degree in order to
compensate for a small $F(d)$ such that $kF(d)\sim 1$. This is for
instance the case for airlines: Short connections go to small airports
while long connections point preferably to big airports, ie. well-connected nodes.
We note that when $F=const.$ we recover the usual preferential
attachment where space is absent \cite{Albert:1999}
and produces a broad distribution of degree with exponent $\gamma=3$.

A first simple case is to consider an exponential function
\begin{align}
F(d)=\mathrm{e}^{-d/d_0}
\end{align}
where $d_0$ sets the typical scale for links. There is therefore
one parameter $\eta=d_0/L$ which governs the behavior of this
model. The degree distribution is of the form \cite{Barthelemy:2003a}
\begin{align}
P(k)\sim k^{-\gamma}f\left(\frac{k}{k_c}\right)
\end{align}
where $\gamma=3$ (the value for the `pure' preferential attachment
model). The cut-off is scaling in dimension $d$ as $k_c\sim (\eta^d)^\nu$ where
the numerical analysis gives $\nu=0.13$ \cite{Barthelemy:2003a}. The
scaling for the average
shortest path is
\begin{align}
\langle\ell\rangle=(N^*)^{\nu'}g\left(\frac{N}{N^*}\right)
\end{align}
where $\nu'\approx 0.3$ and where the scaling function behaves
similarly to the Watts-Strogatz case
\begin{align}
g(x)\sim\begin{cases}
x^{\nu'}\;\;\mathrm{for}\;\;x\ll 1\\
\log x\;\;\mathrm{for}\;\;x\gg 1
\end{cases}
\end{align}
The crossover size $N^*$ depends on the parameter $\eta$ and its
behavior in two extreme cases can be found by the following argument: for $\eta\gg1$ space is irrelevant and
$N^*\sim\mathrm{const.}$, we are always in the small-world regime. In
contrast, when $\eta\ll1$, space is relevant and we expect to have a
crossover from a large to small-world regime. We denote by $a\sim 1/N^{1/d}$ the typical
internode distance and the transition will take place when
\begin{align}
d_0\sim a(N^*)\Rightarrow N^*(\eta\ll 1)\sim \frac{1}{\eta^d}
\end{align}
For $N>N^*$ the network is in the small-world regime and its diameter
is growing as $\langle \ell\rangle \sim \log N$. In the opposite case,
the range $d_0$ is small and the network is much larger: to connect a
typical pair of nodes we need to pass through
$\langle \ell\rangle\sim N^{\nu'}$ points which is found to be smaller
than the lattice-like behavior $N^{1/2}$ for ($d=2$), probably due to
the existence of a few rare long links.

Another simple case was considered where
\begin{equation}
F(d)=d^{-\alpha}
\label{eq:def}
\end{equation}
and which was studied in \cite{Manna:2002,Xulvi:2002,Yook:2002} and
recently rediscussed in \cite{Balister:2018}. The numerical study
presented in \cite{Xulvi:2002} shows that in the one-dimensional case,
for all values of $\alpha$ the average shortest path behaves as
$\log N$. The degree distribution is however different for $\alpha<1$
where it is a power law, while for $\alpha>1$, it is decreasing much
faster and the numerical results in \cite{Xulvi:2002} suggest a
stretched exponential behavior.

\begin{figure}
\begin{tabular}{c}
\includegraphics[angle=0,scale=.40]{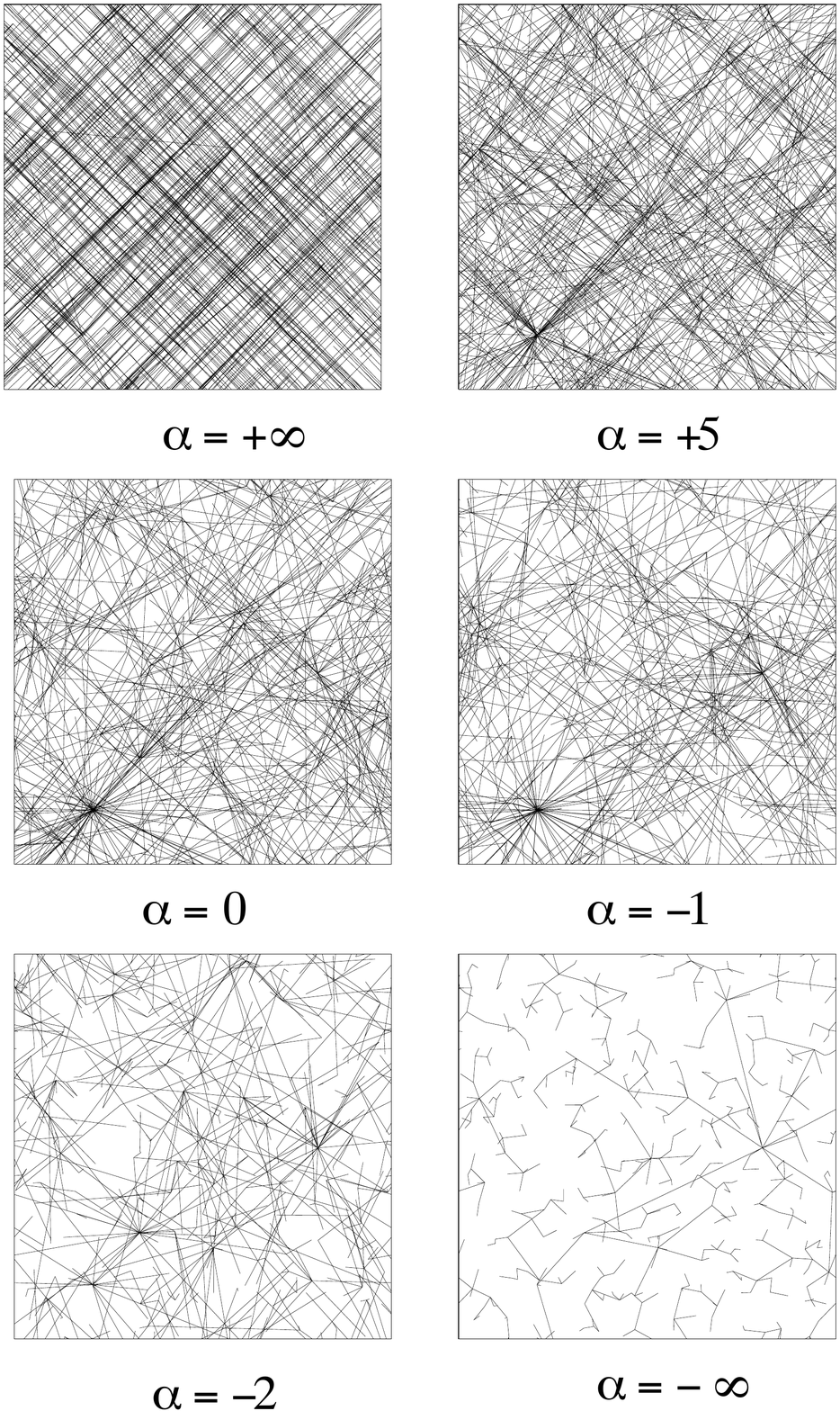}
\end{tabular}
\caption{ Various networks obtained with the rule
  $F(d)=d^{\alpha}$. Figure taken from \cite{Manna:2002}.}
\label{fig:manna}
\end{figure}

In \cite{Manna:2002}, Manna and Sen study the same model but for
various dimensions $d$ and for values of $\alpha$ going from $-\infty$ to
$+\infty$ where the node connects to the farthest and the closest
node, respectively (Fig.~\ref{fig:manna}). Note that in this study \cite{Manna:2002} the convention
on $\alpha$ is different as they chose $F(d)=d^{+\alpha}$ but here and
in the following we stick to the definition Eq.~\eqref{eq:def}.
These authors found a transition for $\alpha=\alpha_c$ and degree
distributions in agreement with the results of
\cite{Xulvi:2002}. They also studied the behavior of the link length
distribution $P(\ell_1)$ and found a power law of the form
\begin{align}
P(\ell_1)\sim \ell_1^{-\delta}
\end{align}
where $\delta=\alpha-d+1$ (which simply results from dimensional
counting). From this result they
define a critical value $\alpha^*=d-1$. Numerically, $\delta$ indeed
behaves linearly up to $\alpha^*$ and then saturates to the value
$\delta_m=d+1$ for $\alpha>\alpha^*$. The following argument can be
given for this value $\delta_m$ in the limit $\alpha\to\infty$. In
this case, a new node always connects to its nearest neighbor and the
probability that the link length $\ell_{t+1}$ of the $t+1^{th}$ node is larger than $\ell_1$ is
given by
\begin{align}
P(\ell_{t+1}>\ell_1)=(1-b\ell_1^d)^t
\end{align}
where $b\ell_1^d$ is the probability to fall in the hypersphere of
radius $\ell_1$ and centered around the new node at time $t+1$. This
implies that the probability to have a link of length
$[\ell_1,\ell_1+\mathrm{d}\ell_1]$ at time $t+1$ is given by
\begin{align}
P(\ell_1; t)\mathrm{d}\ell_1=bdt\ell_1^{d-1}(1-b\ell_1^d)^{t-1}
\end{align}
and for all times up to $t=T$ we obtain
\begin{align}
\nonumber
P(\ell_1)&=bd\ell_1^{d-1}\sum_{t=1}^Tt(1-b\ell_1^d)^{t-1}\\
&\sim \frac{1}{\ell_1^{d+1}}
\end{align}
leading to $\delta_m=d+1$.

This study was complemented by another one by the same
authors \cite{Sen:2003} in the $d=1$ case and where the probability to connect to a node $i$
is given by
\begin{equation}
\Pi_{n\to i}\sim k_i^{\beta}d(n,i,)^{\alpha}
\end{equation}
They studied the phase diagram in the plane
$(\alpha,\beta)$ and found transitions for the degree
distribution as above, and also a change of behavior for the
degree-dependent clustering coefficient $C(k)$. They found numerically
that it behaves as 
\begin{equation}
C(k)\sim k^{-\xi}
\end{equation}
where $\xi$ varies from $0$ to $1$ when $\alpha$ varies from $+\infty$ to
$\-\infty$. We note here that this model was also studied numerically in
\cite{Yook:2002} where a phase diagram in the space $(d,\beta,\alpha)$
is proposed. 

The preferential attachment model with space was recently reconsidered
in \cite{Balister:2018} and 
the existence of the numerically observed transition was mathematically proven. In this 
work, the probability to connect to a node $i$ is given by
\begin{align}
\Pi_{n\to i}=\frac{k_i d(i,n)^{-\alpha}}{Z_n}
\label{eq:pi}
\end{align}
where the normalization constant $Z_n$ is 
\begin{align}
Z_n=\sum_{i'<n}k_{i'}d(i',n)^{-\alpha}
\end{align}
and we refer the interested reader to the paper \cite{Balister:2018}
for more details. We reproduce here their heuristic argument which is in
fact in the spirit of the one given above \cite{Petermann:2006}. It is
interesting to note that all these heuristic arguments are based on a
discussion on the behavior of the normalization $Z_n$, which is also
the case for the demonstration of Kleinberg about navigability in
small-worlds with distributed link lengths~\cite{Kleinberg:2000}. In
the continuous limit the normalization constant reads
\begin{align}
Z_n&\approx\int_{\ell_n}^L r^{-\alpha}r^{d-1}\mathrm{d}r\\
&\approx
\frac
{L^{d-\alpha}-\ell_n^{d-\alpha}}
{d-\alpha}
\end{align}
where $\ell_n$ is the lower cut-off for node $n$ and goes to zero with
increasing density.  If $d>\alpha$ the integral is convergent and we
obtain $Z_n\sim L^{d-\alpha}$ for $\ell_n\to 0$. Spatial fluctuations
are therefore irrelevant in this regime. From the expression for
$\Pi_{n\to i}$ (Eq.~\eqref{eq:pi}), the evolution of the degree
distribution is then described by the following equation
\begin{align}
\frac{dk_i}{dt}&=m\frac{1}{t}\sum_{n=1}^t\frac{k_id(n,i)^{-\alpha}}{Z_n}\\
&\approx \sigma(r_i)\frac{k_i}{t}
\end{align}
where
\begin{align}
 \sigma(r_i)=m\int\frac{|r-r_i|^{-\alpha}}{\sum_jk_j|r-r_j|^{-\alpha}}\rho(r)\mathrm{d}r
\end{align}
This result allowed the authors of \cite{Balister:2018} to show that
the degree distribution is a power law $P(k)\sim k^{-\gamma}$ where
$\gamma=1+1/\sup_r\sigma(r)$. In the case where $\rho(r)=const.$,
$\sigma(r_i)=1/2$ and we recover the usual value $\gamma=3$ for the
preferential attachment. It is interesting to note that space
heterogeneities with a non-uniform density $\rho(r)$ will in general
increase $\sigma(r_i)$ and therefore lower the value of $\gamma$ and
favor the degree inhomogeneity.

In the opposite case $d<\alpha$, the node (denoted by $t$) that arrives at time $t$ will typically connect to
a node $i$ at a distance of order $d(t,i)\sim t^{-1/d}$ (where $t$ is
then the current number of nodes in the network). The typical order of
magnitude of $Z_t$ will then be $Z_t\sim t^{\alpha/d}$. If the node $i$ has a (large) degree $k_i$, the sum $Z_t$ will
be dominated by the $t-i$ term and we will have $Z_t\approx
k_id(t,i)^{-\alpha}$ which implies $d(t,i)\sim
k_i^{1/\alpha}t^{-1/d}$. Since nodes are located at random, the probability
to connect to $i$ is then of order $d(t,i)^d$: $\Pi_{t\to i}\sim
d(t,i)^d\sim k_i^{d/\alpha}/t$. The degree evolution equation is then
of the form \cite{Balister:2018}
\begin{align}
\frac{dk_i}{dt}=C\frac{k_i^{d/\alpha}}{t}
\end{align}
which leads for the degree distribution to a stretched exponential of
the form $P(k)\sim \mathrm{exp}-k^{1-d/\alpha}$. Exact results
\cite{Balister:2018} confirm this heuristic argument and the existence
of a sharp transition for $\alpha_c=d$. This transition can be seen in
various quantities such as the degree distribution that goes from
apower law to a stretched exponential when $\alpha$ crosses $d$. In
other words, for $\alpha>d$ spatial constraints are important while
for $\alpha<d$ the growth is mostly governed by the preferential
attachment.

We note here that the average path length does not display here the same
behavior as models A and B above that undergoes a transition from
a small-world to a large world regime characterized by
$N^{1/d}$. Indeed in both phases here, the average shortest path is
`small': it varies as $\log N$ for $\alpha>d$ and as
$\log N/\log\log N$ for $\alpha<d$ \cite{Balister:2018}.

\subsection{Greedy models: cost-benefit analysis}

We will discuss in the next section models of networks defined by the
optimization of a single quantity that depends on the global structure
of the network. In contrast, we consider here the growth of networks
where nodes are added one by one, located at random and connected to
the network in an optimal way.  In general, if we denote by $i$ the
new node, it will connect to the node $j$ (that belongs to the
existing network) such that the quantity
\begin{align}
Z(i,j)=\mathrm{Benefit}(i,j)-\mathrm{Cost}(i,j)
\label{eq:cba0}
\end{align}
is maximum. This quantity represents the balance between the cost of
constructing the link $i-j$ and the benefit that it will create. The
optimization is therefore not global -- the resulting network does not
necessarily optimize some quantity -- but is local. In this respect
these models can be qualified as `greedy' as they rely on a local
optimal choice, but with no guarantee that the system as a whole will
reach a global optimum.

This type of models was proposed by transportation scientists
\cite{Black:1971} and more recently by computer scientists
\cite{Fabrikant:2002} for describing the Internet growth and which
predicts correctly a scale-free degree distribution as observed
empirically (see for example \cite{Pastor:2003} and references
therein). In this model, the functional maximized at each
node addition reads
\begin{align}
Z(i,j)=-g(j)-\lambda C(i,j)
\end{align}
and if we allow only one link per new node the resulting network is a
tree (if the initial conditions are tree-like). The quantity $g(j)$ is
in general a measure of the `centrality' of the node $j$ such as the
average number of hops to other nodes, or to a given central node,
etc. and that we wish to be small (and the benefit $-g$ large). The quantity $C(i,j)$ is a cost
function, in general proportional to the euclidean distance
$d(i,j)$. The quantity $\lambda$ controls the relative importance of
centrality versus distance \cite{Fabrikant:2002}:
\begin{itemize}
\item{} For $\lambda\gg 1$, only the cost
(distance) is minimized and each new node will connect to the nearest
node in the growing cluster; the resulting network will be akin to a
dynamical version of the minimum spanning tree (MST). 
\item{} For $\lambda\simeq 0$, cost has no importance and the new node will
connect to the most central node, producing in general some sort of
star graph (i.e. the complete bipartite graph $K_{N-1,1}$). 
\end{itemize}
Since the euclidean distance between nodes is typically of order
$1/\sqrt{N}$ the value that distinguishes large from small values of
$\lambda$ is ${\cal O}(\sqrt{N})$. Fabrikant et
al. \cite{Fabrikant:2002} also showed that if $\lambda$ has some
intermediate values, we obtain various networks, for example
\cite{Fabrikant:2002}:
\begin{itemize}
\item{} if $1\ll \lambda\ll \sqrt{N}$, we obtain a graph with a degree
  distribution of the form $P(k)\sim k^{-\gamma}$ where $\gamma$
  depends on $\lambda$.
\item{}  if $\lambda\sim N^{1/3}$ there is an exact bound $\gamma\geq 1/6$ and
numerical results seem to indicate that $0.6\leq\gamma\leq 0.9$
\end{itemize}

More generally, we will observe a large variety of networks according
to the choice of the functions $g(j)$ and $C(i,j)$ and we will
consider here essentially an example of cost-benefit models
\cite{Black:1971,Louf:2013b} which depends on a single scale and
produces a family of networks which range from the star-graph to the
minimum spanning tree and which are characterised by a continuously
varying exponent \cite{Louf:2013b}. More precisely, we assume that all
nodes are distributed uniformly in the plane and are labelled by a
quantity $P_i$. For a rail network, each node corresponds to a city
and has a population $P_i$ which is assumed to be distributed
according to a power-law with exponent $1+\mu$ with $\mu\approx 1.0$
(see \cite{Louf:2013b} for further details and discussions). The edges
are added sequentially to the graph -- as a result of a cost-benefit
analysis -- until all the nodes are connected. For the sake of
simplicity, we describe here the growth of trees which allows to focus
on the emergence of large-scale structures due to the cost-benefit
ingredient alone.  Motivated by transportation networks, the cost is
chosen to be proportional to the euclidean distance $d(i,j)$ between
$i$ and $j$
\begin{align}
\label{eq:cost}
C(i,j)= \kappa d(i,j)
\end{align}
where $\kappa$ represents the cost of a link per unit length. Benefits
are more difficult to assess, and here also motivated by
transportation networks, we assume that the benefit is proportional to
the expected traffic between nodes. We then use the common and simple
assumption of the so-called gravity law (see \cite{Erlander:1990} and
references therein) to estimate the traffic and we end up with the
following expression for the cost-benefit budget (up to irrelevant
factors) \cite{Louf:2013b}
\begin{align}
\label{eq:R0}
Z_{ij} = K\frac{P_i P_j}{d(i,j)^{\;a-1}} - \: \beta d(i,j)
\end{align}
where $\beta$ represents the relative importance of the cost with
regards to the benefits and which is the main control
parameter here (the constants are here $a>1$, $K>0$).

If we denote by $\overline{P}$ the average population and we note that
the  typical inter-city distance is of order $1/\sqrt{\rho}$ ($\rho=N/L^2$ 
denotes the city density, and $L$ is the typical linear
size of the whole system), it is easy to see that the two 
terms of Eq.~\eqref{eq:R0} are of the
same order for $\beta=\beta^*$ defined as
\begin{align}
\label{eq:beta*}
\beta^* \propto \overline{P}^2 \rho^{a/2}
\end{align}
From Eq.~\eqref{eq:beta*} we can guess the existence of two different
regimes depending on the value of $\beta$:
\begin{itemize}
\item $\beta \ll \beta^*$ the cost term is negligible compared to the
  benefits term. Each connected city has its own influence zone
  depending on its population and the new cities will tend to connect
  to the most influent city. In the case where $a\approx 1$, every
  city connects to the most populated cities and we obtain a star
  graph constituted of one single hub connected to all other cities.
\item $\beta \gg \beta^*$ the benefits term is negligible compared to
  the cost term. All new cities will connect sequentially to their
  nearest neighbour. If we select the node $i$ such that the length of
  the link $i-j$ is the smallest, the algorithm is then equivalent to
  Prim's algorithm~\cite{Prim:1957}, and the resulting graph is a
  minimum spanning tree (MST).
\end{itemize}

\begin{figure}[!h]
\centering
\includegraphics[width=0.5\textwidth]{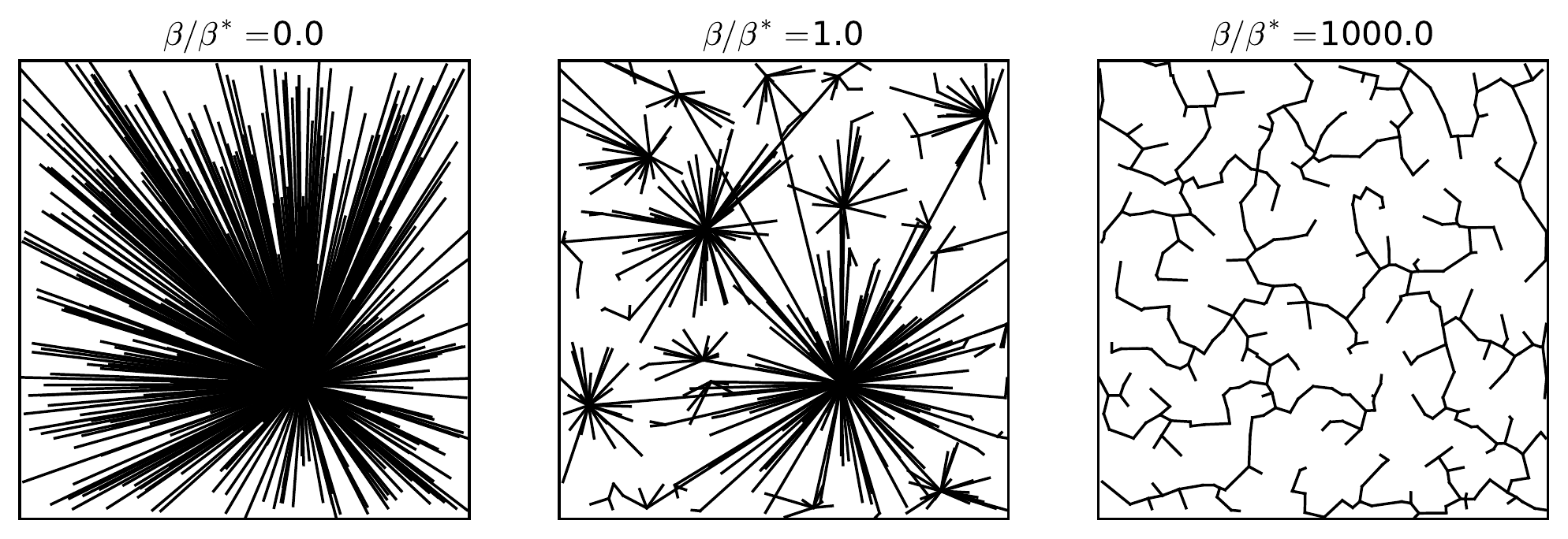}
\caption{Graphs obtained with the cost-benefit algorithm for the 
same set of cities (nodes) for three different values of 
$\beta^*$ ($a=1.1$, $\mu=1.1$, $400$ cities). On the 
left panel, we have a star graph where the most populated 
node is the hub and on the right panel, we recover the minimum spanning tree. Figure taken from \cite{Louf:2013b}.
  \label{fig:plot_graphs}}
\end{figure}

Fig.~\ref{fig:plot_graphs} shows three graphs obtained for the same
set of cities for three different values of $\beta/\beta^*$ ($a=1.1$,
$\mu=1.1$) confirming the discussion above about the two extreme
regimes.

For $\beta \sim \beta^*$ we observe a different type of graph, which
suggests the existence of a crossover between the star-graph and the
MST. This graph is reminiscent of the hub-and-spoke structure that has
been used to describe the interactions between city
pairs~\cite{Okelly:1996,Okelly:1998}, but it is important to note that
this structure emerges naturally in the system and does not result
from a global optimization. Also, we note that the graph corresponding
to the intermediate regime $\beta \approx \beta^*$ (see
Fig.~\ref{fig:plot_graphs}) exhibits a particular structure
corresponding to a hierarchical organization observed in many complex
networks~\cite{Sales:2007} and which can be characterized
quantitatively. In particular,
the average euclidean distance $\overline{d}$ between the different
hierarchical levels decreases with the level rank (for details, see \cite{Louf:2013b}).

\begin{figure}[!h]
\centering
\includegraphics[width=0.40\textwidth]{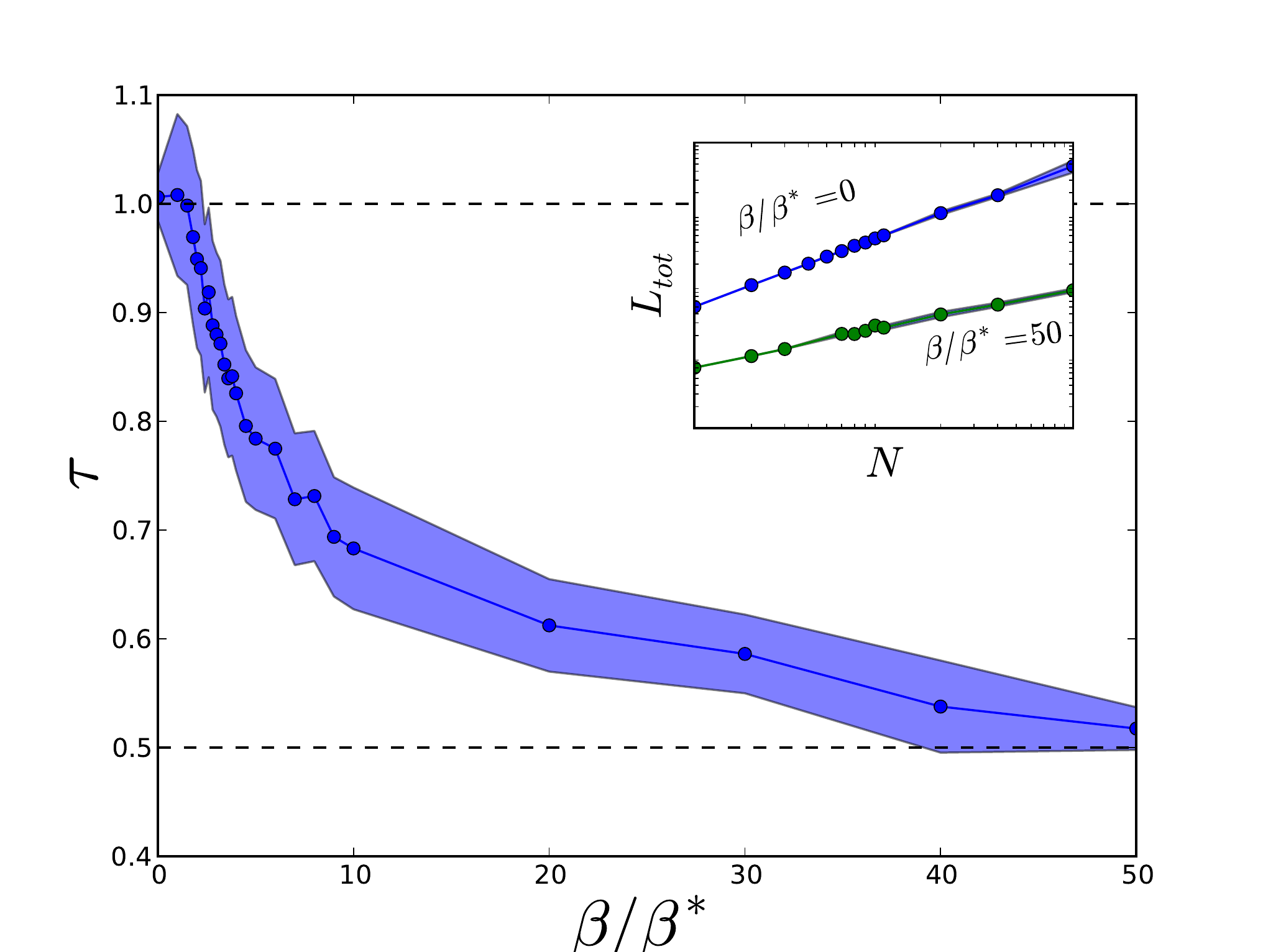} 
\caption{Exponent $\tau$ versus $\beta$. For $\beta\ll \beta^*$ we
  recover the star-graph exponent $\tau=1$ and for the other extreme
  $\beta\gg\beta^*$ we recover the MST exponent $\tau=1/2$. In the
  intermediate range, we observe a continuously varying exponent
  suggesting a non-trivial structure. The shaded area represents the
  standard deviation of $\tau$. Inset: In
  order to illustrate how we determined the value of $\tau$, we
  represent $L_{tot}$ versus $N$ for two different values of
  $\beta$. The power law fit of these curves gives $\tau$. Figure
  taken from \cite{Louf:2013b}.}
\label{fig:Ltot_vs_beta}
\end{figure}

An important difference between the star-graph and the MST lies
in the scaling of the total length of the graph with its number of
nodes. Indeed, in the case of the star-graph, all the nodes are
connected to the same node and the typical edge length is $L$, the
typical size of the system the nodes are enclosed in. We thus obtain
\begin{align}
\label{eq:Ltot_star}
L_{tot} \sim L\; N
\end{align}
On the other hand, for the MST each node is connected roughly to its
nearest neighbour at a distance typically given by
$\ell_1\sim L/\sqrt{N}$, leading to
\begin{align}
\label{eq:Ltot_MST}
L_{tot} \sim L\; \sqrt{N}
\end{align}
More generally, we expect a scaling of the form 
\begin{align}
L_{tot}\sim N^\tau
\end{align}
and on Fig.~\ref{fig:Ltot_vs_beta} we show the variation of this
exponent $\tau$ versus $\beta$. For $\beta=0$ we have $\tau=1.0$ and
we recover the behavior $L_{tot} \propto N$ typical of a star
graph. In the limit $\beta \gg \beta^*$ we also recover the scaling
$L_{tot} \propto \sqrt{N}$, typical of a MST. For intermediate values,
we observe an exponent which varies continuously in the range
$[0.5,1.0]$.

\begin{figure}[!h]
\centering
\includegraphics[width=0.25\textwidth]{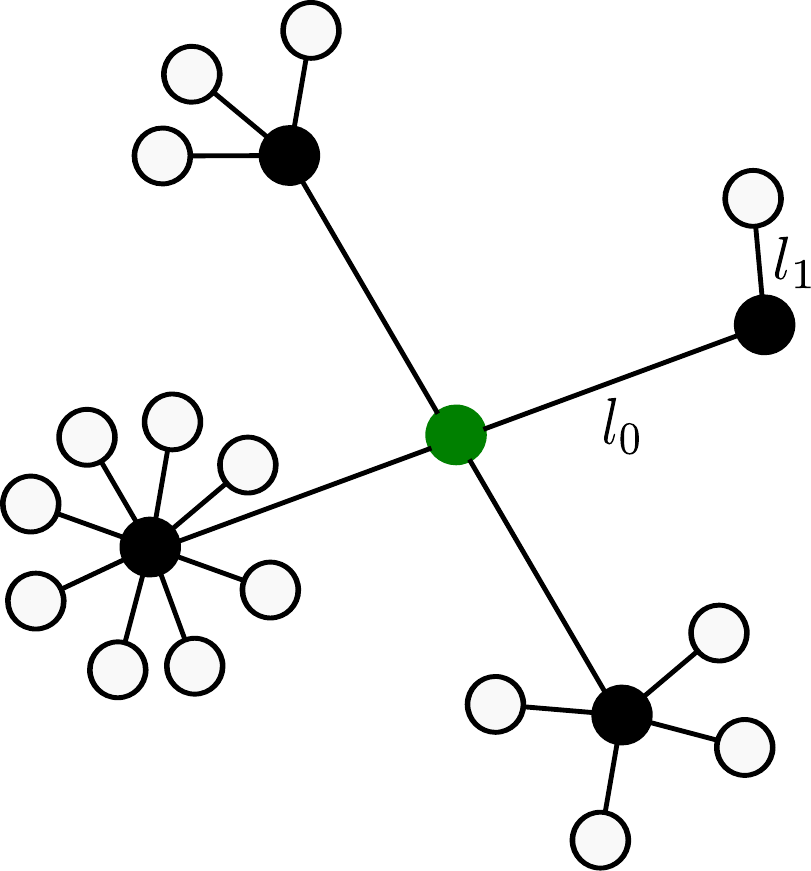}
\caption{A schematic representation of the hierarchical fractal 
network used as a toy model. Figure taken from \cite{Louf:2013b}.
  \label{fig:fractal_network}}
\end{figure}

We can understand this crossover behavior as a consequence of the
hierarchical structure of these graphs with a simple toy model
\cite{Louf:2013b}. More precisely, we consider the fractal tree
depicted on Fig.~\ref{fig:fractal_network} constructed recursively as
a tree of connectivity $z$ (in this figure only $3$ levels are shown).
For this model, the distance between the levels $n$ and $n+1$ is given
by $\ell_n=\ell_0 b^n$ where $b\in [0,1]$ is the scaling factor. This
scaling factor would then correspond to the average distance between
successive hierarchical level $\overline{d}$ and thus decreases with
the depth in the hierarchy. For a
regular tree, each node at the level $n$ is connected to $z$ nodes at
the level $n+1$ which implies that $N_n = z^n$ where $z>0$ is an
integer. A simple calculation shows that in the limit $z^g \gg 1$, the
total length of the graph with $g$ levels scales as
\begin{align}
L_{tot} \sim N^{\frac{\ln(b)}{\ln(z)}+1}
\end{align}
where $\frac{\ln(b)}{\ln(z)} +1 \leq 1$ because $b \leq 1$ and
$z>1$. This simple model thus provides a simple mechanism where the
exponent for $L_{tot}$ varies continuously and depends on the scaling
factor $b$. The parameter z can be easily determined from the average
degree of the network, and the parameter $b$ can be related to our
model by measuring the decrease of the mean distance $\overline{d}$
between different levels of the hierarchy \cite{Louf:2013b}. This toy
model thus provides a simplified picture of the graphs in the
intermediate regime $\beta \simeq \beta^*$ and exhibits their key
features in this regime: the hub structure reminiscent
of the star graph and nodes connected to each hub forming
geographically distinct regions and organized in a hierarchical fashion.

Finally, it is interesting to note that empirical networks seem to
be in the range $\beta\sim\beta^*$ suggesting that spatial hierarchy
and other features not discussed here (see \cite{Louf:2013b})
are relevant for real-world networks. It seems plausible that
the general cost-benefit framework could be applied to the modelling
of systems besides transportation networks: it captures the
fundamental features of spatial network while being versatile enough
to model the growth of a great diversity of systems shaped in part by
spatial constraints. From this point of view, it would therefore be 
very interesting to understand the effect of other benefit
functions and the conditions for observing crossovers and
transitions.

\section{Localization transition}

In general, the structure of a graph has naturally an effect on the pattern of
shortest paths on it and we will illustrate this phenomenon on simple
examples. We will first discuss a simple structure formed by a ring
and radial spokes present with a probability $p$ and we will compute
the average shortest path. Beyond the average shortest path
we will also characterize the effect of structural changes with the help of
the betweenness centrality (BC), a simple proxy for traffic. In
particular, we will discuss two examples where a variation in the
structure induces dramatic changes in the pattern of congested nodes
or `bottlenecks' characterized by a large BC value.

\subsection{Hub-and-spokes structure}

\begin{figure}[h!]
\centering
\includegraphics[width=0.50\textwidth]{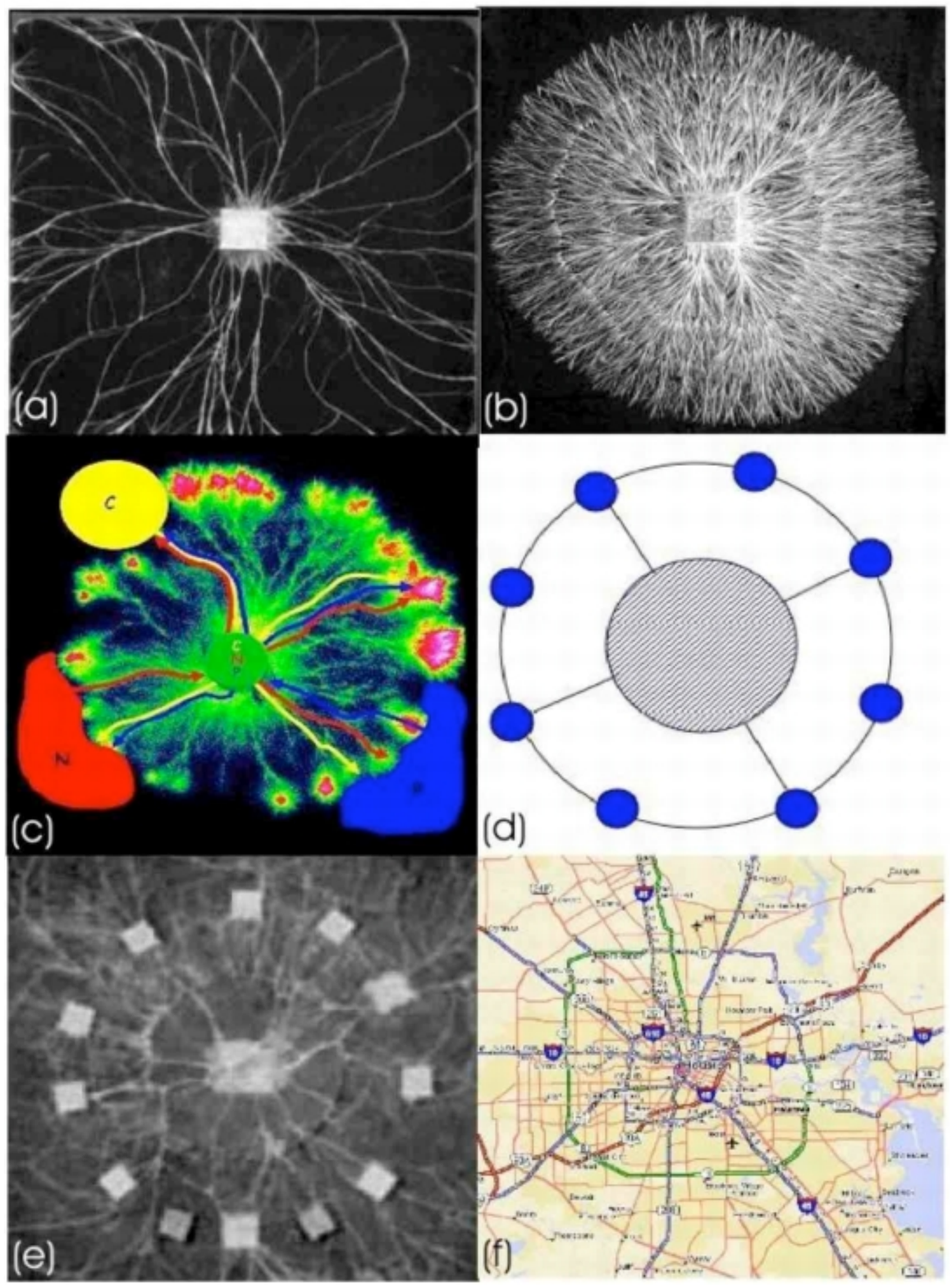}
\caption{ Examples of hub-and-spoke structures with rings. (a-c,e):
  Typical fungi networks, in (c) a schematic representation of the
  nutrient flow is shown. (d) The model studied in
  \cite{Dorogovtsev:2000a,Ashton:2005,Jarrett:2006} with spokes radiating
  from a hub. (f) Road network in Houston showing an inner hub with a
  complicated structure. Figure taken from \cite{Jarrett:2006}. }
\label{fig:ashton}
\end{figure}

In many real-world cases the `pure' hub-and-spoke structure is not
present and we observe in general a ring structure around a
complicated core or an effective hub (see some examples in
Fig.~\ref{fig:ashton}). We will however use this coarse-grained
structure and present an interesting discussion proposed in
\cite{Ashton:2005,Jarrett:2006} about centralization versus
decentralization from the perspective of the average shortest path.

\begin{figure}[h!]
\centering
\includegraphics[width=0.45\textwidth]{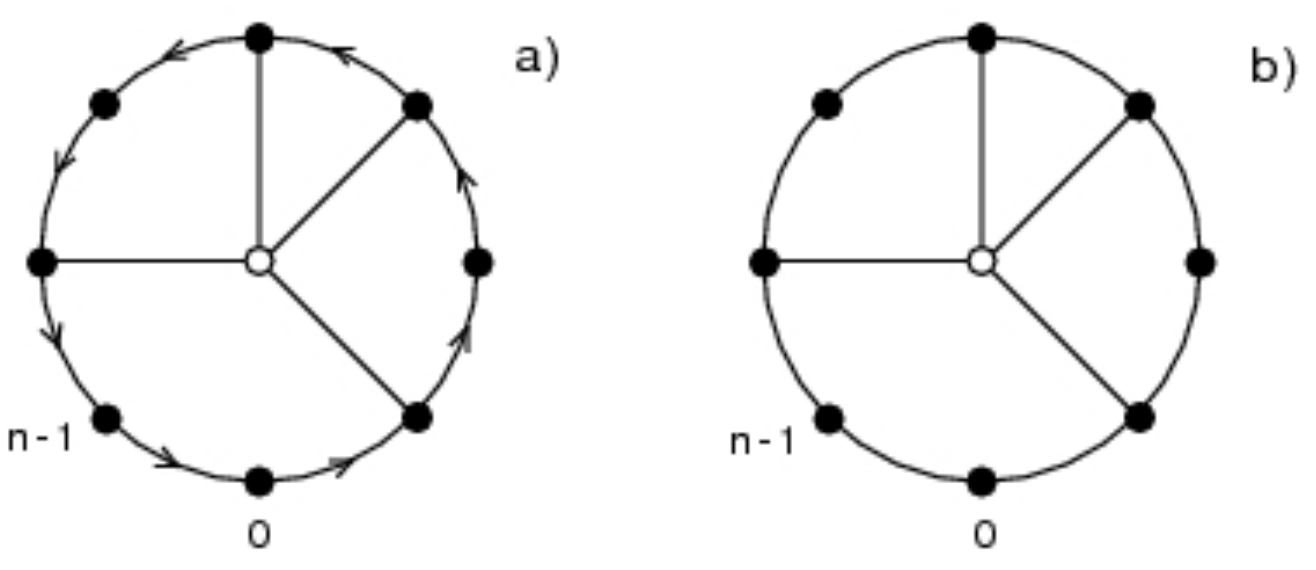}
\caption{ Models proposed and discussed in \cite{Dorogovtsev:2000a}
  and studied with congestion in \cite{Ashton:2005,Jarrett:2006}. A
  central site is connected to a site on a ring with probability
  $p$. In (a) all the links on the ring are directed and in (b) these
  links are not directed. Figure taken from
  \cite{Dorogovtsev:2000a}. }
\label{fig:ashton2}
\end{figure}

The main ingredient in these models is the competition between the centralized
organization with shortest paths going through a single central hub and
decentralized paths going along a ring and avoiding the central hub (in
particular in presence of congestion).  A simple model of hub-and-spoke structure
together with a ring was proposed in \cite{Dorogovtsev:2000a} where
$N$ nodes are on a circle and a hub is located at the center of
the circle (see Fig.~\ref{fig:ashton2}). Each radial link -- a spoke --
is present with probability $p$. 

We first consider this model and reproduce the
results obtained in \cite{Dorogovtsev:2000a} for the average shortest
path $\langle\ell(p)\rangle$ and its distribution $P(\ell)$. We
discuss the simpler case where the loop is oriented as shown in
Fig.~\ref{fig:ashton2}(a) (for undirected links calculations are more
involved but results are similar). The central point is connected
with undirected links of weight $1/2$ added with probability $p$. This
amounts to connect random pairs of nodes by undirected links of length
$1$. Obviously $\langle\ell(p=0)\rangle\sim N$ while
$\langle\ell(p=1)\rangle=1$ showing that we have for this simple model a
crossover between a large and a small-world behavior when $p$ is
varied. In the case of directed links, the result for the shortest
path distribution is \cite{Dorogovtsev:2000a}
\begin{align}
P(\ell)=\frac{1}{N-1}\left[1+(\ell-1)p+\ell(N-1-\ell)p^2\right](1-p)^{\ell-1}
\end{align}
and the expression for the average
shortest path $\langle\ell\rangle=\sum_{\ell=1}^N\ell P(\ell)$ is
\begin{align}
 \langle\ell\rangle=\frac{1}{N-1}\left[\frac{2-p}{p}N-\frac{3}{p^2}+\frac{2}{p}+\frac{(1-p)^N}{p}(N-2+\frac{3}{p})\right]
\end{align}
We can easily check that this expression is consistent with the two limiting cases
\begin{align}
&\text{For}\;\; p\to 0:\;\; P(\ell)\to 1/(N-1)\Rightarrow \langle\ell\rangle=N/2\\
&\text{For}\;\; p\to 1:\;\;P(\ell)\to \delta_{\ell,1}\Rightarrow
  \langle\ell\rangle=1
\end{align}
As expected this simple model thus displays a crossover from a large to small-world when $p$
increases. At small $p$ most shortest paths go along the ring, while
for $p$ close to 1, most shortest paths go through the central hub.

An interesting observation made in \cite{Ashton:2005} is that if we
add a cost $c$ each time a path goes through the central hub, we
could expect some sort of transition between a decentralized regime
where it is less costly to stay on the peripheral ring to a
centralized regime where the cost is not enough to divert paths
from the central hub. The cost could in general depend on how busy the
center is and could therefore grow with the number of connections to
the hub. In the case of a constant cost $c$ (and in the directed
case), we can estimate the shortest path distribution ($N$ is here the
number of nodes on the ring) and the result is \cite{Ashton:2005}
\begin{align}
P(\ell)=
\begin{cases}
\frac{1}{N-1}\;\;\; &{\rm for}\;\; \ell\leq c\\
\frac{1}{N-1}[1+b_1p+b_2p^2](1-p)^{\ell-c-1}\;\;\;
&{\rm for}\;\; \ell >c
\end{cases}
\end{align}
where $b_1=\ell-c-1$ and $b_2=(N-1-\ell)(\ell-c)$. For paths of length
$\ell\leq c$, there is no point to go through the central hub. In the
opposite case, when $\ell >c$, we recover a distribution similar to
the $c=0$ case of \cite{Dorogovtsev:2000a}.  The average shortest path
is now
\begin{align}
\nonumber
\langle\ell\rangle&=\frac{(1-p)^{N-c}[3+(N-2-c)p]}{p^2(N-1)}\\
&+\frac{p[2-2c+2N-(c-1)(c-N)p]-3}{p^2(N-1)}+\frac{c(c-1)}{2(N-1)}
\label{eq:lmin}
\end{align}
We can consider the continuous limit ($p\to 0$, $N\to\infty$, and $z\equiv\ell/N$
and $\rho=pN$ fixed), the average shortest path is then a function of
these parameters $(\rho,c,N)$. This analysis allows to show that there is an
optimal number of connections to the central hub in order to minimize
the average shortest path. For example for a cost increasing linearly
as $c=k\rho$ the optimal value is $\rho^*\sim \sqrt{N/k}$ which
corresponds to the simple condition $pc(\rho)\sim 1$ (for more details
see \cite{Ashton:2005,Barthelemy:2011}). This study \cite{Ashton:2005} was generalized in \cite{Jarrett:2006}
to the case of a more complicated cost function such as
$c(\rho)=C\rho+B\rho^2+A\rho^3$ where the authors observe different
behaviors and a phase transition according to the values of the
coefficients $A$, $B$, and $C$. 

These studies on a simple toy model show how congestion could have an
important impact on the a priori optimal hub-and-spoke structure and
favorizes the transport along a ring. From a more general perspective
it would indeed be interesting to observe the emergence of rings -- as
observed in real-world examples -- without imposing it a priori.

\subsection{A loop and branches toy model}

We saw in the previous section how the density of spokes can influence the
pattern of shortest paths. In order to go beyond the average shortest
path, we will measure the variations of the shortest path pattern with
the help of the betweenness centrality (BC). The BC is a measure of
the importance of a node (or an edge) and is a very simple proxy for traffic on the network. More
precisely, the betweenness centrality $g(i$) of node $i$ is defined as \cite{Freeman:1977,Newman:2001b,Goh:2001,Barthelemy:2003b,Barthelemy:2003c}.
\begin{align}
g(i)=\frac{1}{\cal N}\sum_{s\neq t}\frac{\sigma_{st}(i)}{\sigma_{st}}
\label{eq:bc}
\end{align}
where $\sigma_{st}$ is the number of shortest paths going from $s$ to
$t$ and $\sigma_{st}(i)$ is the number of shortest paths going from
$s$ to $t$ through the node $i$ (the normalization ${\cal N}$ is
usually chosen as ${\cal N}=(N-1)(N-2)$ which counts the number of
pairs and ensures that $g(i)\in [0,1]$). This quantity $g(i$) thus
characterizes the importance of the node $i$ in the organization of
flows in the network (note that with this definition, the betweenness
centrality of terminal nodes is zero, and that a similar definition
can be given for the BC of edges). The bottlenecks in the graph are therefore the
nodes with a large BC. These nodes are critical for shortest paths and
their study allows us to highlight important structural changes.

\begin{figure}[!]
\centering
\includegraphics[width=0.35\textwidth]{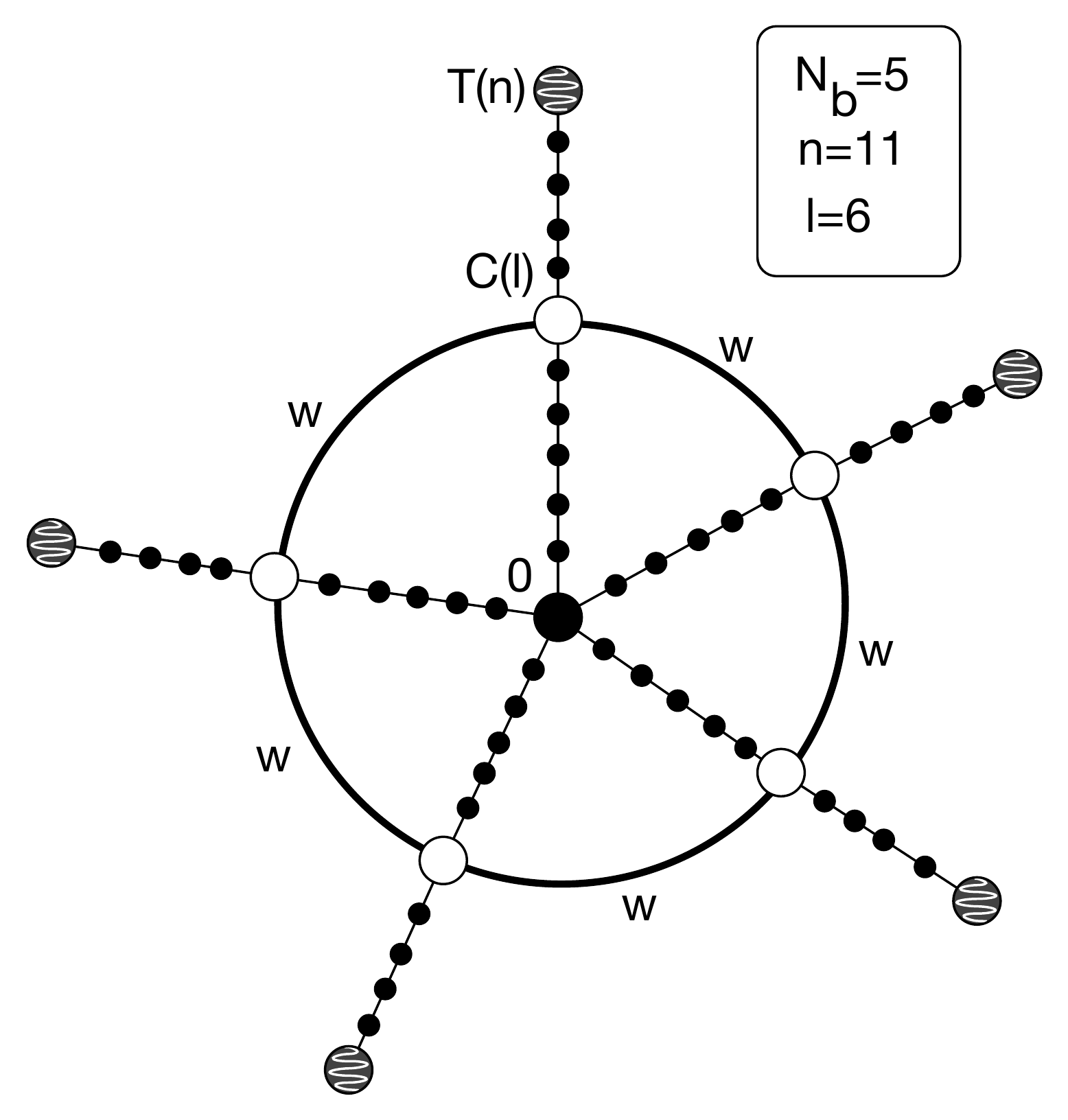}
\caption{ Representation of the toy model discussed
  here. The number of branches is here $N_b=5$, the number of nodes on
  each branch is $n=11$ and the loop is located at a distance $\ell=6$
  from the center $0$. The node $C$ is at the intersection of a branch
and the loop and $T$ is the terminal node of a branch. Figure taken
from \cite{Lion:2017}.}
\label{fig:toymodel0}
\end{figure}

We expect in general that the pattern of shortest paths on a network
will be strongly affected by the link properties: if the links are
weighted by the time needed to cross them, the quickest path will
result from an interplay of the topology and the weight structure. In
order to understand if we can observe transitions in the pattern of
quickest paths on a network and motivated by the crossover described
in the previous section, we will consider a simple model that
incorporates (i) a center, (ii) radial links and (iii) a loop.  Such a
simple toy model was considered in \cite{Lion:2017} where the network
is constructed on a star network composed of $N_b$ branches, each
branch being composed of $n$ nodes. We then add a loop at a distance
$\ell$ from the center (see Fig.~\ref{fig:toymodel0} for a sketch of
this graph). We also consider the general case where links are
weighted and have a weight equal to one for radial links, and a weight
$w$ for links on the loop (between two consecutive branches) -- we can
then recover the purely topological case by taking $w=1$. The quantity
$w$ can be seen as the time spent on the segment and the
weighted shortest path is then the quickest path. We compute
the BC using weighted shortest paths (the `quickest' paths) and this generalization allows us
to discuss the impact of different velocities on a street network. 

\begin{figure}[h!]
\centering
\includegraphics[scale=0.30]{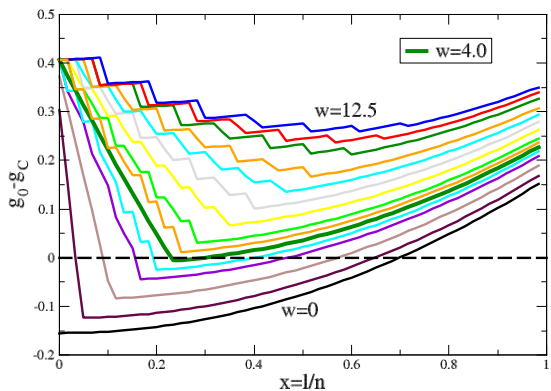}
\caption{ Difference $\delta g(\ell)=g_0-g_C$ versus $\ell/n$ for $N_b$ and $n$ fixed and
  for different values of $w$ in the range $[0,12.5]$. For values less
  than a threshold ($w_c\approx 4$ here)  there is a minimum that is
  negative. Figure from \cite{Lion:2017}}
\label{fig:deltag0}
\end{figure}

Within this simple toy model, we will discuss under which conditions
the loop is more central than the `origin' at the center. As above,
the structure of the network is fixed and we focus on transitions in
the spatial pattern of shortest paths when the weight $w$ is
modified. Intuitively, for very large $w$, it is always less costly to
avoid the loop, while for $w\to 0$, the loop is very advantageous.
The two main quantities of interest are therefore the centrality at
the center denoted by $g_0(\ell,n,w)$ and the centrality at the
intersection $C$ of the branch and the loop, denoted by
$g_C(\ell,n,w)$. The interest of this toy model lies in the fact that
we can estimate analytically these quantities and we will give here
the main results and refer to \cite{Lion:2017} for more details. We
want to understand here if the origin has always a larger BC than the
loop, or if for some values of the parameters, this order can be
inverted. We will then compute the difference $\delta g=g_0-g_C$ and
study under which condition it can be negative. We plot this quantity
versus $\ell$ for different values of $w$ and we observe the result
shown in Fig.~\ref{fig:deltag0}
This result shows that for $w$ sufficiently small, $\delta g$ can be
negative. This demonstrates the existence of a threshold value $w_c$
such that at $w=w_c$ the minimum is $\min_\ell\delta g=0$. 
For $w<w_c$, the minimum of $\delta g$ is negative and we can define
an optimal value $\ell_{opt}$ which corresponds to this smallest value
of $\delta g$. The quantity $\ell_{opt}$ thus gives the position of
loop that maximizes the difference between the BC of the loop and
the center. 

In \cite{Lion:2017}, all the derivations for the quantities $w_c$ and
$\ell_{opt}$ can be found, and we restrict ourselves here to
hand-waving arguments. In order to determine $\ell_{opt}$, we note
that when $\ell$ is small, most paths connecting nodes from different branches
will go through $0$ and we expect $\delta g>0$. When $\ell$ is
increasing more paths will go through the loop and will increase the
value of $g_C$. However, when $\ell$ is too large, paths connecting
the (large) fraction of nodes located on the lower branches will go
through $0$ again. In order to get a sufficient condition on
$\ell_{opt}$, we consider the path between the node $C$ on the branch
`1' and the corresponding node $C'$ on the furthest branch
$(N_b-1)/2$ (for $N_b$ odd). The optimal value for $\ell_{opt}$ is then such that the
cost of the path from $C$ to $C'$ through $0$ which is given by $2\ell$ is
equal to the cost on the loop equal to $w(N_b-1)/2$. This
leads to the result
\begin{align}
\ell_{opt}\approx \frac{w(N_b-1)}{4}
\end{align}
(which is the exact result).

The threshold quantity $w_c$ is obtained by imposing that the minimum
of $\delta g(\ell=\ell_{opt})$ is equal to zero, but 
we can understand the scaling for $w_c$ with the simple following
argument. Indeed, a necessary condition on $w$ is that $\ell_{opt}$
must be less than $n$ which leads to the condition
\begin{align}
w<w_c\sim 4\frac{n}{N_b}
\end{align}
(the scaling $n/N_b$ is exact but the prefactor's value seems to
be closer to $2/3$ \cite{Lion:2017}).

If we come back to the `topological' case where all weights are equal
to $1$, these results on this simple toy model show that the loop can
be more central than the origin if $w_c>1$ which implies that
$n\gg N_b$. It thus suggests that the number and the spatial extension
of radial branches are crucial ingredients that control the
existence of central loops. If the extension $n$ of the network is large
compared to the number of radial branches, $w_c$ can be larger than
one $w_c > 1$ and central loops can be observed for $w = 1$. In
ordered systems -– such as lattices -– the effective number of
branches is too large leading to a very small $w_c$ and therefore
prohibits the appearance of central loops in the `topological’ case
($w = 1$). In real-world planar graphs where randomness is present,
the absence of some links can lead to a small number of `effective’
radial branches which in the framework of the toy model implies a
large value of $w_c$ and therefore a large probability to observe central
loops.

\subsection{Tuning the edge density}

We considered previously a simple toy model and we now discuss a more
general one where randomness is present. In particular, we want to
understand the BC distribution $P(g)$ and the spatial pattern of
bottlenecks -- the high BC nodes -- when the density (of nodes or links) increases. For spatial
networks that are trees the BC is very large and belongs in the range
$[N,N^2]$: for terminal nodes and their neighbors, we have a $g\sim N$
behavior while for central nodes we expect the scaling $g\sim N^2$. In
general networks are not trees and contain loops, and their presence
creates alternate paths producing a wealth of nodes with a low BC. We
thus expect the emergence of a low BC regime with increasingly sharp
cut-offs with increasing density.

In order to discuss and confirm this idea, we discuss 
the simple model of planar graph proposed in \cite{Kirkley:2018}.
We distribute uniformly $N$ nodes in the 2d plane and generate
the Delaunay triangulation on this set. We then remove at random links
so that the edge density defined by
\begin{equation}\label{eqn:edge_density}
\rho_e=\frac{E}{ E_{DT}}, 
\end{equation}
reaches a desired value. In this expression, we normalize the number of edges $E$ by its value $E_{DT}$ for
the Delaunay triangulation. For a maximally planar graph (i.e. one in which
no more edges can be added without violating the planarity
constraint), we have $E_{DT} \approx 3N$, which implies that the 
density captures a quantity proportional to the average degree. The
density $\rho_e$ thus varies from $\approx 1/3$ for the MST to $1$ for the
DT~\cite{Lee:1980}, and allows us to monitor changes in these graphs.

\begin{figure}[h!]
 \includegraphics[width=0.45\textwidth]{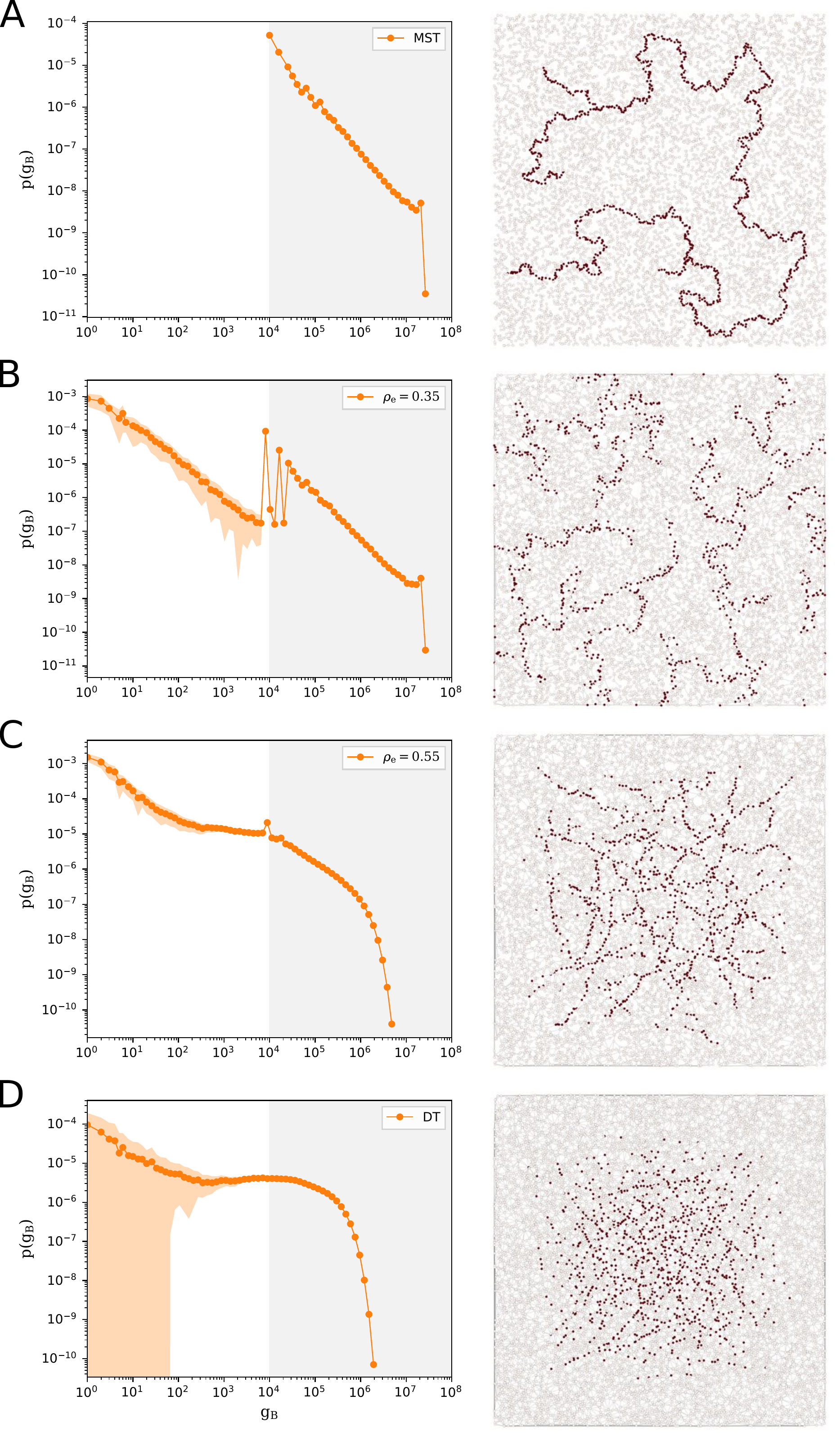}
 \caption{{\bf Effect of edge-density $\rho_e$ on the betweenness}.
   (Left panels) BC distribution for the MST and for the model
   obtained by pruning the Delaunay triangulation for various values
   of the density (obtained for $N =10^{4}$ and a hundred
   realizations). As a guide to the eye, the `tree-like' region
   (shaded) is distinguished from the `loop-like' region below
   $N$. (Right panels) Generated network with the corresponding edge
   density $\rho_e$. Shown in red are the nodes in the $90^{th}$
   percentile and above in terms of their BC value. Figure taken from
   \cite{Kirkley:2018}.}
\label{fig:fig3}
\end{figure}

We show results obtained with this simple model (and for comparison,
we also show the results for the minimum spanning tree) in
Fig.~\ref{fig:fig3}. On the left panels, we show the BC distribution
and on the right panels the corresponding  network with the nodes
having the largest BC ($90^{th}$ percentile).
As expected, the BC distribution for the MST (Fig.~\ref{fig:fig3}A) is
peaked at $N$ and is bounded by $N^2/2$ (which corresponds here to the
interval $[10^4, 10^8]$). In this interval, the BC distribution
follows a form close to a calculation on a Cayley tree (see
\cite{Kirkley:2018} for more details). For increasing density
$\rho_e$, we have loops in the graph which induces the bimodal shape
of the BC distribution (Fig.~\ref{fig:fig3}B-D). The low BC in the
range $[1,N^2]$ results from these loops that allow to bypass some of
the high BC nodes. For larger values of $\rho_e$ the distribution gets
progressively more homogeneous while keeping the peak around $N$ even
as we approach the limiting case of the DT (Fig.~\ref{fig:fig3}D).

\begin{figure}[t!]
\centering 
\includegraphics[width=\linewidth]{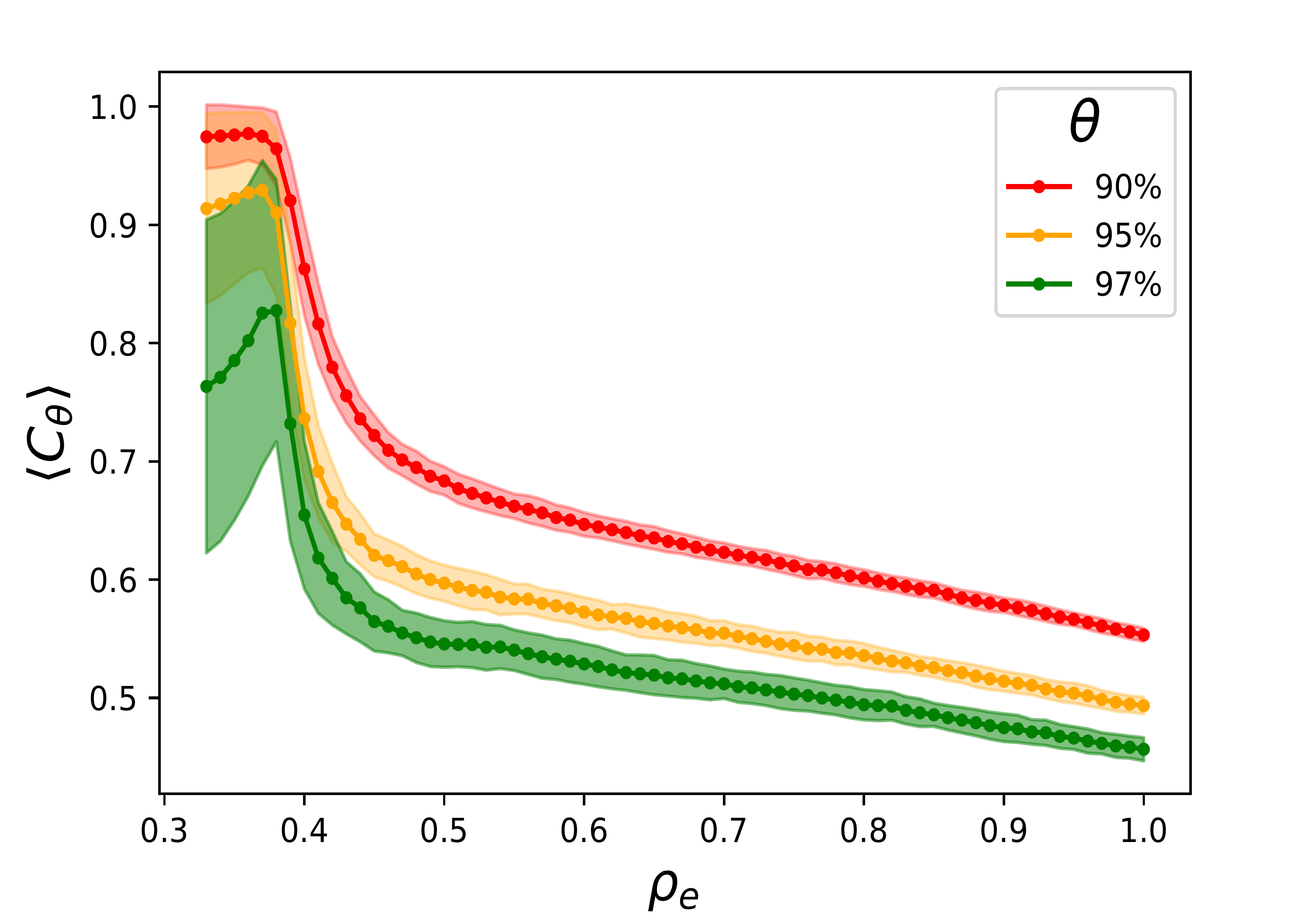} 
\caption{{\bf Behavior of high BC nodes.} The clustering index
  $\langle C_\theta \rangle$ (for the $\theta$ percentile of high BC
  nodes) decreases for denser networks, capturing the tendency of the
  nodes to be increasingly clustered around their center of
  mass. Figure from \cite{Kirkley:2018}.}
\label{fig:3_clustering_metrics}
\end{figure}

These simulations also reveals an interesting spatial behavior. Nodes
with a large BC are highlighted in the right-hand panels of
Fig.~\ref{fig:fig3} and display a change in their spatial pattern with
increasing $\rho_e$. In the MST case, the high BC nodes span the whole
network, and for increasing density, we observe a localization trend
where the high BC nodes cluster together around the barycenter. This
suggests some sort of topological transition (or crossover) towards a
spatial regime characterized by a clustering of the BC around the
barycenter and which corresponds roughly to the regular lattice case
where the BC is a smoothly decreasing function with the distance to
the barycenter. We can quantify more precisely this transition by
measuring various indicators (see \cite{Kirkley:2018}), and we will
here discuss the clustering of BC nodes above the $\theta^{th}$
percentile. We compute their spread around their center of mass
normalized by the average distance of nodes to the
center of mass
\begin{align}
C_{\theta}=\frac{N}{N_\theta}\frac{\sum_{i=1}^{N_\theta}|x_i-x_{cm}|}
{\sum_{i=1}^{N}|x_i-x_{cm}|}
\label{eq:cluster}
\end{align}
where the average position is
$x_{\textrm{cm}}=\sum_{i=1}^{N_\theta}x_i$, and
where $N_\theta$ is the number of high betweenness nodes at percentile
$\theta$ (the quantities $\{x_i\}$ specify the coordinates of
nodes). The clustering index (Eq.~\eqref{eq:cluster}) thus quantifies
the extent of clustering of the high BC nodes relative to the rest of
the nodes in the network, with increased clustering resulting in low
values of $C_{\theta}$. In Fig.~\ref{fig:3_clustering_metrics} we plot
the average quantity $\langle C_{\theta} \rangle$ for various values
of $\theta$ ($90$, $95$, and $97$) versus $\rho_e$ and we observe a
clear decrease with the density ($\langle \ldots \rangle$ indicates
averaging over realizations). Indeed the decrease is approximately by
a factor of two from the MST to the DT, confirming the spatial
clustering of the nodes to be a robust effect. We note here that in
order to confirm if this is a transition or a crossover, a further
analysis of finite size effect is needed. All the metrics point to
the fact that the spatial location of a node has little relevance to
its BC in a sparse network, whereas it assumes increasing importance
for denser networks. This is confirmed by plotting the rescaled
average BC of nodes as a function of the distance $r$ from the
barycenter (see \cite{Kirkley:2018}): for low values of $\rho_e$ there
appears no distance dependence of the nodes, whereas for
$\rho_e > 0.4$, a clear $r$ dependence emerges with the curves
converging to the form seen for dense random geometric
graphs as calculated in~\cite{Giles:2015}.

Interestingly emough, real-world cases are considered in
\cite{Kirkley:2018} (street network of 97 world cities, and the
evolution of central Paris), and are shown to display a behavior that
is well described by this simple model (we refer the interested reader
to the paper \cite{Kirkley:2018} for more details and discussion about
real-world cases). This simple model seems thus able to describe the
`localization' transition of bottlenecks with increasing densities
(Fig.~\ref{fig:fig3}), and can be thought of as a proxy for the
evolution of a urban road network as it experiences refinements in infrastructure
with increased connectivity.

\section{Optimal networks}

An important class of networks are obtained by optimizing a functional
of a graph. A simple and well-known example is the minimum spanning
tree (MST) that minimizes the total length of a tree connecting
together a given set of points. Even if most existing spatial networks
in the real-world seem not to result from a global optimization but
rather evolve through a progressive growth process, the interest in
optimal networks lies in the fact that they constitute interesting
benchmarks to compare actual networks with. For example, the
comparison of a real-world network with the MST constructed over the
same set of nodes indicates how far we are from the minimum cost
possible, an important information for applications.

Here we will focus on a small set of examples and will leave aside the
important litterature about practical applications of graph
optimization. An important example in transport network applications
is the hub-and-spoke structure (see for example \cite{Okelly:1998} and
references therein) where direct connections are replaced with fewer
connections to hubs which form a network at a larger scale. The
hub-and-spoke structure reduces the network costs, centralizes the
handling and sorting, and allows carriers to take advantage of scale
economies through consolidation of flows. Instead we will focus on
transition aspects and we will essentially discuss a family of optimal
networks that comprises the MST and the star graph. We will also
discuss new results about the effect of congestion on some optimal
networks and finally, how the existence of fluctuations can induce the
emergence of loops in optimal networks.

\subsection{A family of optimal trees}

\begin{figure}[h!]
\centering
\includegraphics[width=0.5\textwidth]{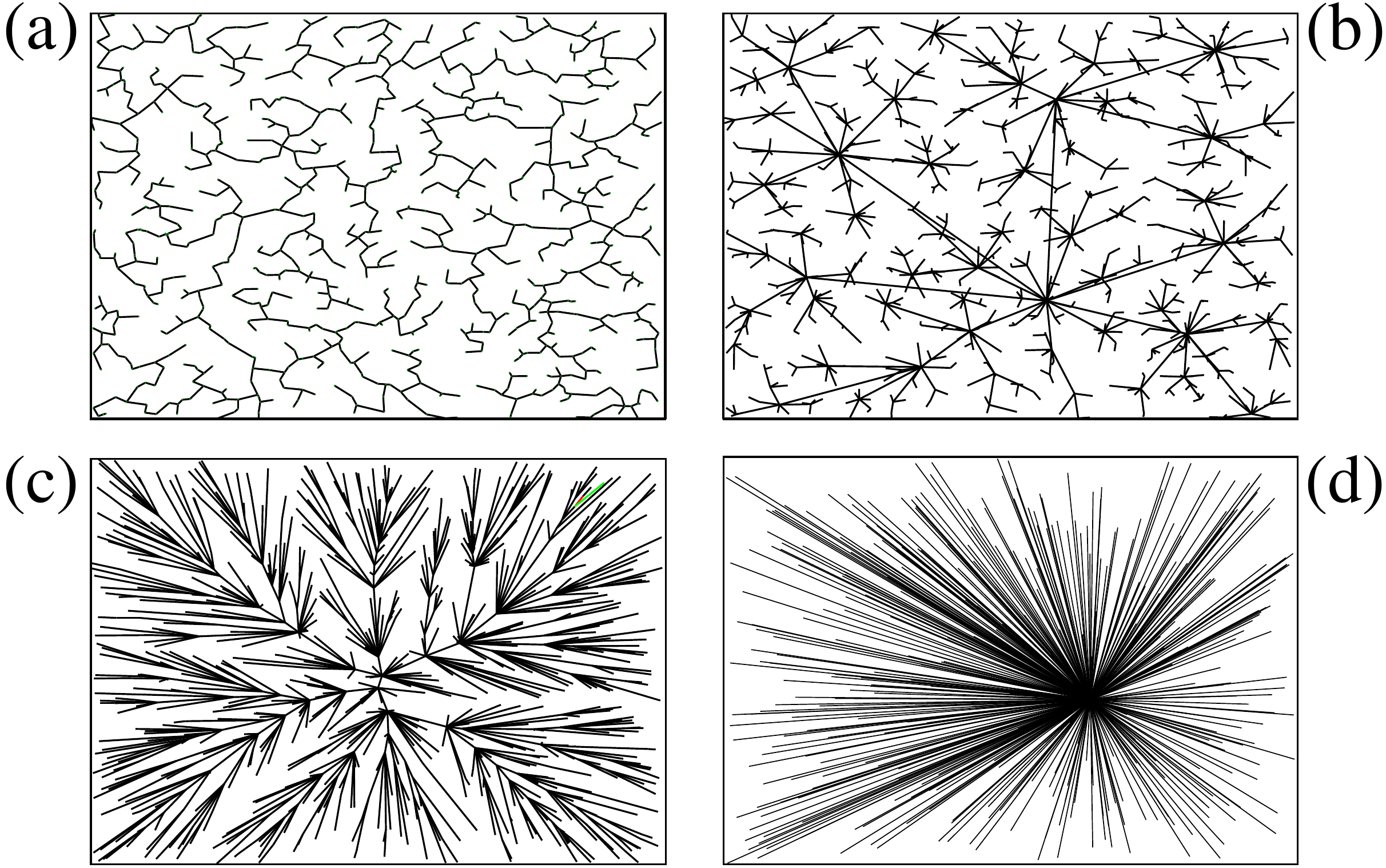}
\caption{ Different spanning trees obtained for different values of
$(\mu,\nu)$ in Eq.~\eqref{eq:munu} for the same set of $N=1000$
nodes. (a) Minimum spanning tree obtained for $(\mu,\nu)=(0,1)$: in
this case the total distance is minimized. (b) Optimal traffic tree
obtained for $(\mu,\nu)=(1/2,1/2)$. In this case we have an interplay
between centralization and minimum distance resulting in local
hubs. (c) Minimum euclidean distance tree obtained for
$(\mu,\nu)=(1,1)$. In this case centrality dominates over distance and a
`star' structure emerges with a few dominant hubs. (d) Optimal
betweenneess centrality tree obtained for $(\mu,\nu)=(1,0)$. In this
case we obtain the shortest path tree which has one star hub (for the
sake of clarity, we omitted some links in this last figure). Figure
taken from \cite{Barthelemy:2006}.}
\label{fig:examples}
\end{figure}

The minimum spanning tree is defined as the graph which minimizes the
total cost given by a sum over all edges belonging to this tree. The
weight of a link is in general a simple local function and does not
depend on the other links. Another simple example of an optimal tree
is the star graph (all nodes are connected to a single hub) that
minimizes the average shortest path (if one moves a link from this
configuration, it is easy to see that the average shortest path can
only grow). At this point we could ask if there is a general framework
for describing these different optimal trees and if there is a
relation between them. In \cite{Barthelemy:2006} it was suggested that
these optimal trees can be viewed as particular cases of the
minimization of the following quantity
\begin{align}
E_{\mu\nu}=\sum_{e\in{\cal T}} g(e)^{\mu}d(e)^{\nu}
\label{eq:munu}
\end{align}
where $g(e)$ is the BC of edge $e$ and $d(e)$ is the length of this
edge, and the sum is over all edges on the tree ${\cal T}$. The
exponents $\mu$ and $\nu$ control the relative importance of distance
against topology as measured by centrality. Fig.~\ref{fig:examples}
displays examples of spanning trees obtained for different values of
$(\mu,\nu)$.

\begin{figure}
\centerline{
\includegraphics*[angle=-90,width=0.4\textwidth]{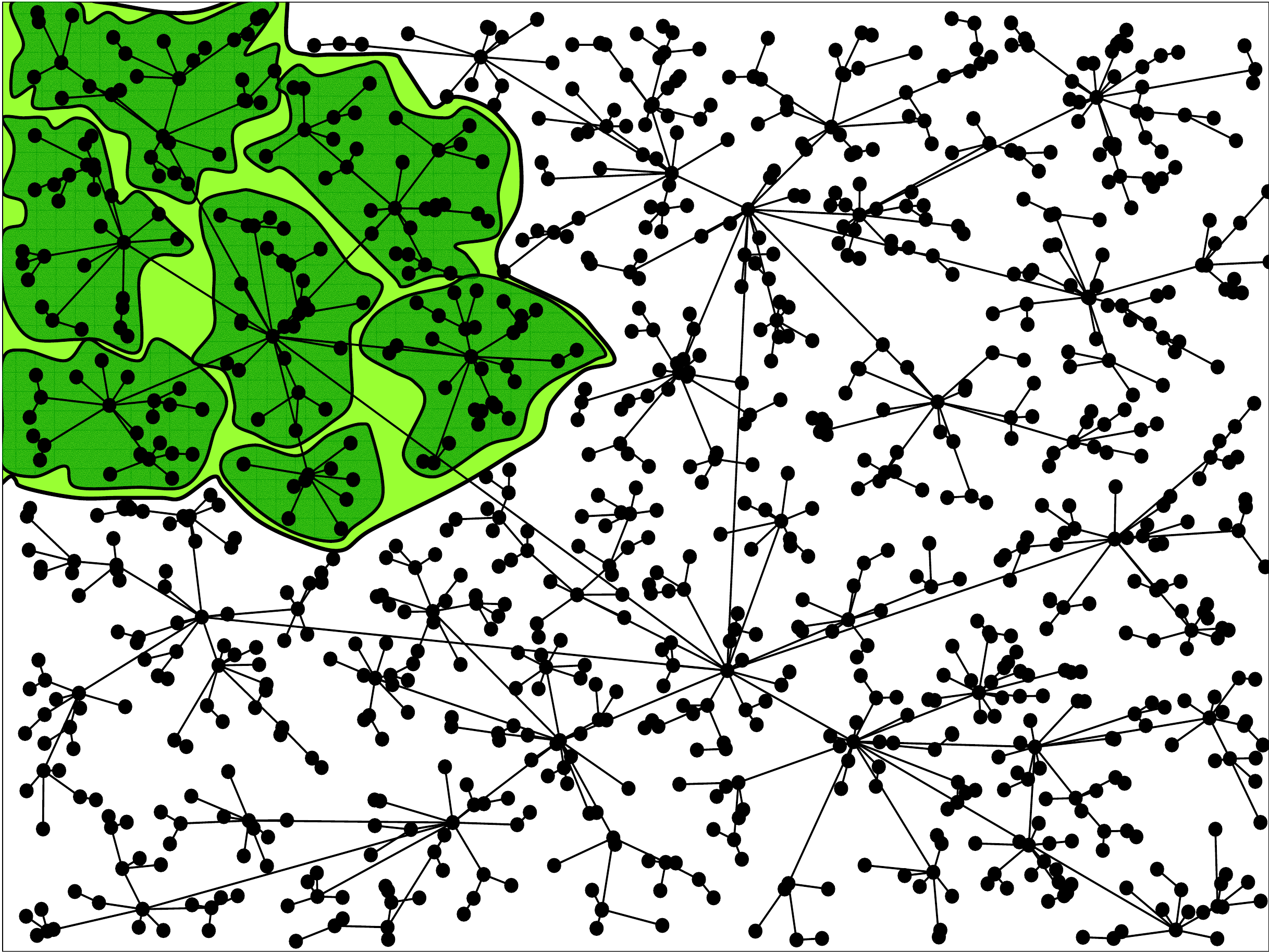}
}
\caption{Hierarchical organization emerging for the optimal traffic
tree $(\mu,\nu)=(1/2,1/2)$ (for $N=1000$ nodes). Longer links lead to
regional hubs which in turn connect to smaller hubs distributing
traffic in smaller regions. Figure taken from \cite{Barthelemy:2006}.}
\label{fig:regions}
\end{figure}

For $(\mu,\nu)=(0,1)$ we recover the euclidean minimum spanning tree
(see Fig.~(\ref{fig:examples}a)). In the case $(\mu,\nu)=(1,0)$, the
energy \eqref{eq:munu} is proportional to the average betweenness
centrality which is also proportional to the average shortest path
$\sum_e b_e\propto \langle\ell\rangle$ (see for example
\cite{Barthelemy:2018}).  The tree $(1,0)$ shown in
Fig.~(\ref{fig:examples}d) is thus the shortest path tree (SPT) with
an arbitrary `star-like' hub (a small non zero value of $\nu$ would
select as the star the closest node to the gravity center). For
intermediate cases such as $(\mu,\nu)=(1/2,1/2)$ we obtain the
`optimal traffic tree' (OTT) (see Fig.~(\ref{fig:examples}b)) which
displays an interesting interplay between distance and shortest path
minimization (see \cite{Barthelemy:2006}). The spatial properties of
the OTT are also remarkable, in particular it displays
(Fig.~\ref{fig:regions}) a hierarchical spatial organization
where long links connect regional hubs, that, in turn are
connected to sub-regional hubs, etc. This hierarchical structure can
be probed by measuring the average euclidean distance between
nodes belonging to the cluster obtained by deleting recursively the longest link.
\begin{figure}
\includegraphics[width=0.4\textwidth]{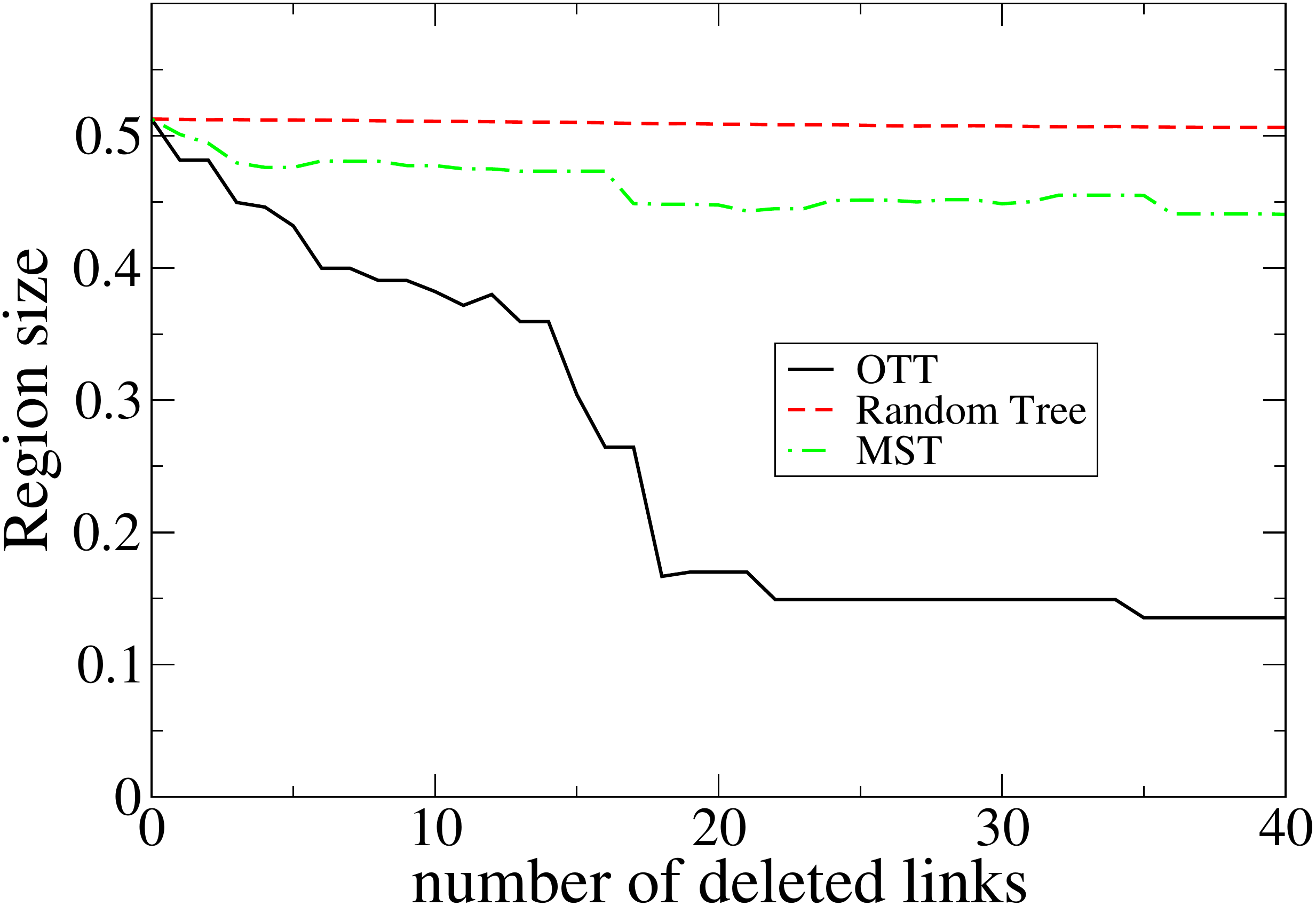}
\caption{ Average euclidean size of the largest cluster remaining
after deleting links ranked according to their length (in decreasing
order) obtained for one typical configuration of size $N=1000$ of
different networks. The decrease observed for the OTT is consistent
with a hierarchical spatial organization as it is visually evident
from Fig.~\ref{fig:regions}. Figure taken from \cite{Barthelemy:2006}.}
\label{fig:ave_d}
\end{figure}
For the OTT, we observe a decrease of the region size, demonstrating
that longer links connect smaller regions, a feature absent in
non-hierarchical networks such as the MST, the shortest path tree or
the random tree (see Fig.~\ref{fig:ave_d}).

Finally, for $(\mu,\nu)=(1,1)$, the energy is proportional to the
average shortest weighted path (with weights equal to euclidean
distance (see Fig.~(\ref{fig:examples}c)) and in this case centrality
dominates over distance and a `star' structure emerges with a few
dominant hubs.

The minimization of Eq.~\eqref{eq:munu} thus provides a natural
interpolation between the MST and the SPT, a problem which was
addressed in previous studies~\cite{Khuller:1995}. The degree
distribution for all cases considered above is not broad, possibly as a consequence
of spatial constraints. We expect that when we vary continuously the parameters
$\mu$ and $\nu$, there will be transitions between these different
structures and a complete inspection of the plane $(\mu,\nu)$ would be
very interesting.

\subsection{Congestion-induced transition}

Congestion will naturally act on the shortest paths organization. In
the case where the network is fixed and has one hub, the study
\cite{Ashton:2005} demonstrated the existence of an optimal number of
connections that minimizes the total traveling time.  In another work
\cite{Guimera:2002b}, the authors studied the structure of networks
optimizing the average search cost both in the presence and in the
absence of congestion. They demonstrated the existence
of a transition from a centralized structure for low congestion
situations to a more decentralized topology for highly congested
cases. Here, we will show that a transition from centralized to a
decentralized network is actually present for the lowest
non-trivial form of the cost which contains interactions between
users, a way to describe congestion.

\begin{figure}
\includegraphics[width=0.5\textwidth]{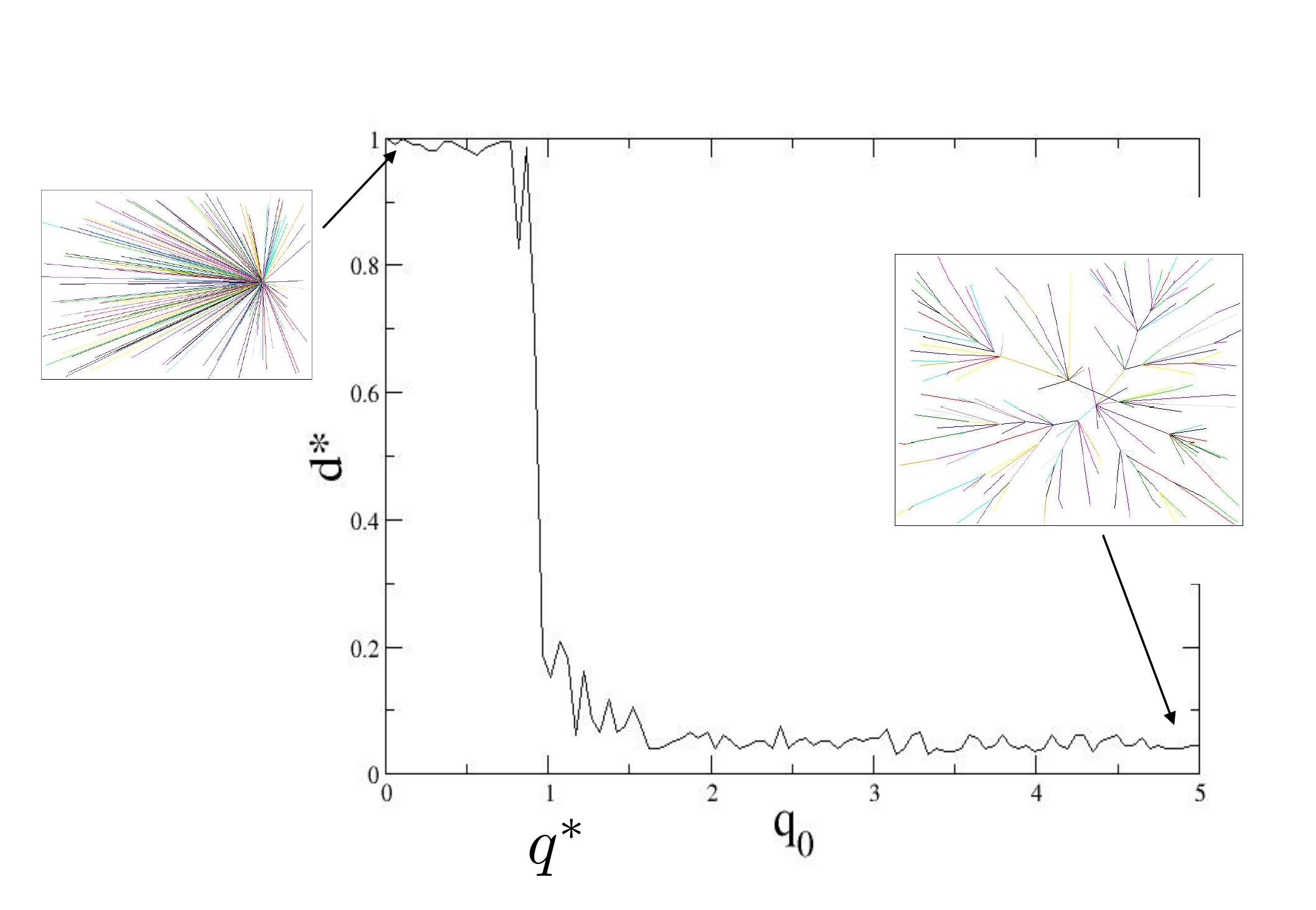}
\caption{ Degree dominance versus capacity. For a capacity of order
  $q_0=q^*\approx 1$ there is a sharp transition between a one star
  network to a multipolar structure (simulation obtained for $N=1,000$
nodes).}
\label{fig:dstar}
\end{figure}

\begin{figure}
\includegraphics[width=0.35\textwidth]{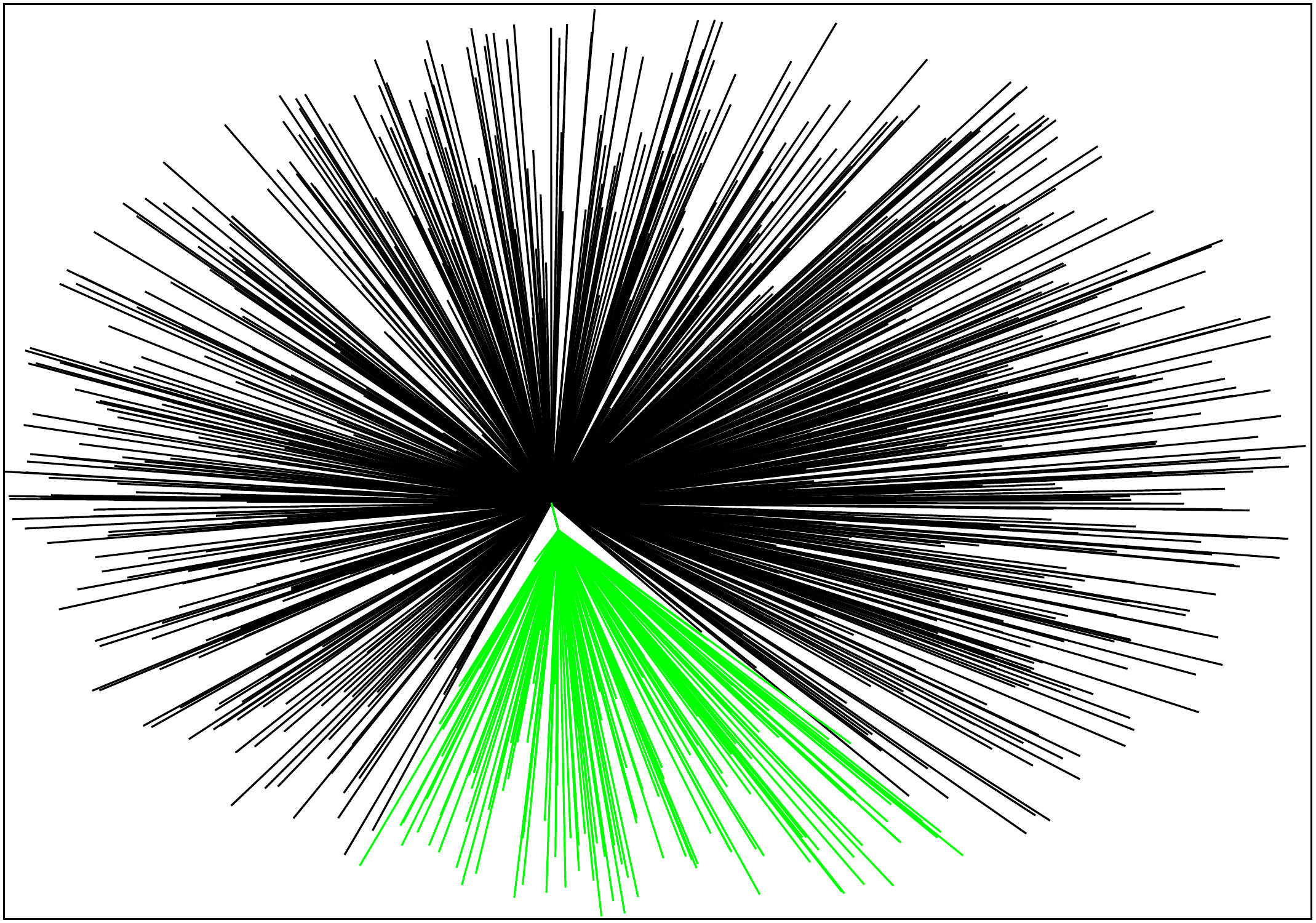}
\caption{ The network observed numerically right after the
  transition (here $N=1000$ nodes and $q_0=0.95$).}
\label{fig:3a}
\end{figure}

\begin{figure}
\includegraphics[width=0.4\textwidth]{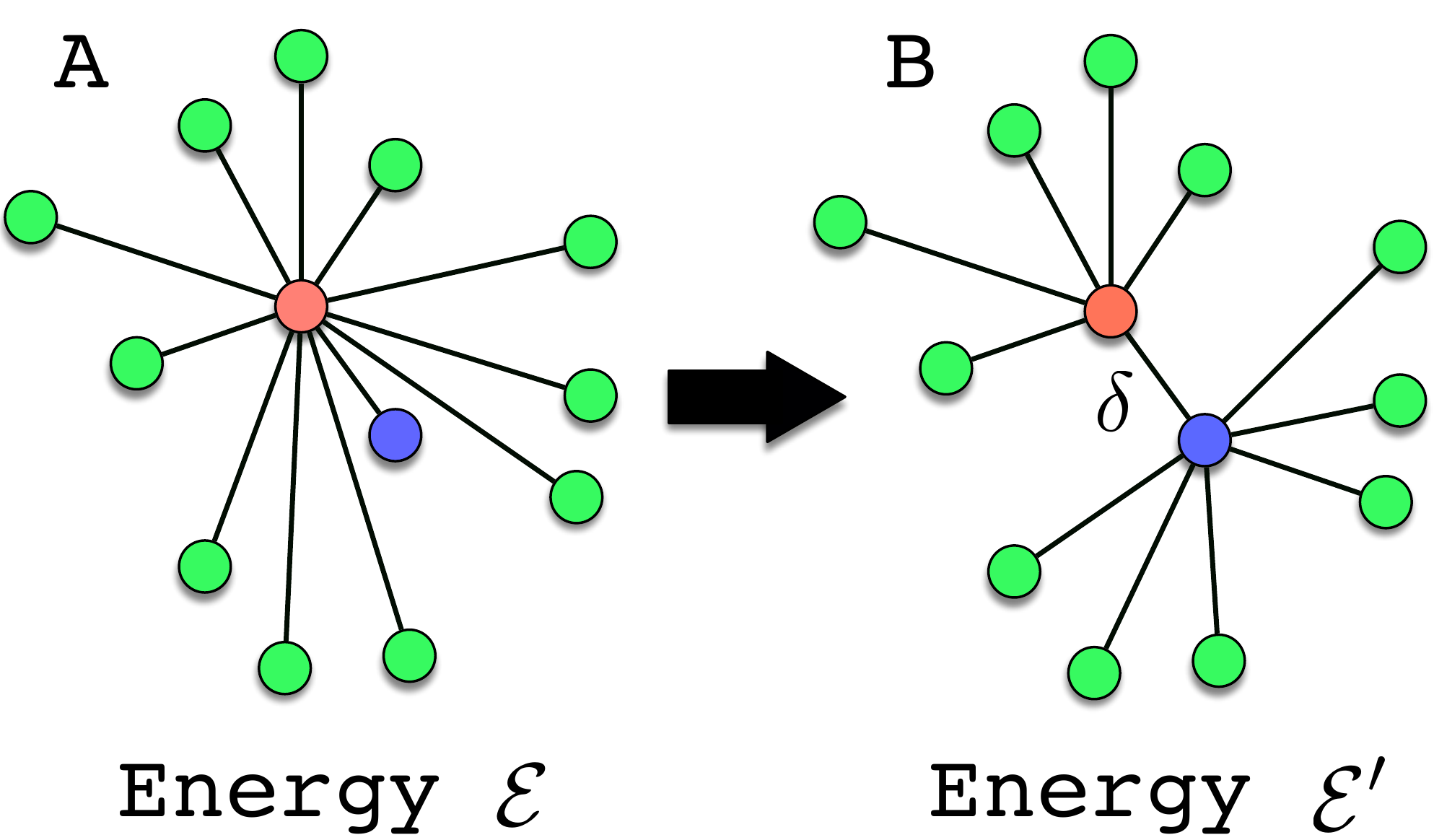}
\caption{Simplified picture used for the theoretical calculation of
  $q^*$. On the left, the initial star network has an energy
  ${\cal E}$. On the right, the nearest site of the first hub becomes
  a hub itself. The energy of this configuration is ${\cal E}'$. }
\label{fig:3b}
\end{figure}

\begin{figure}
\includegraphics[width=0.25\textwidth]{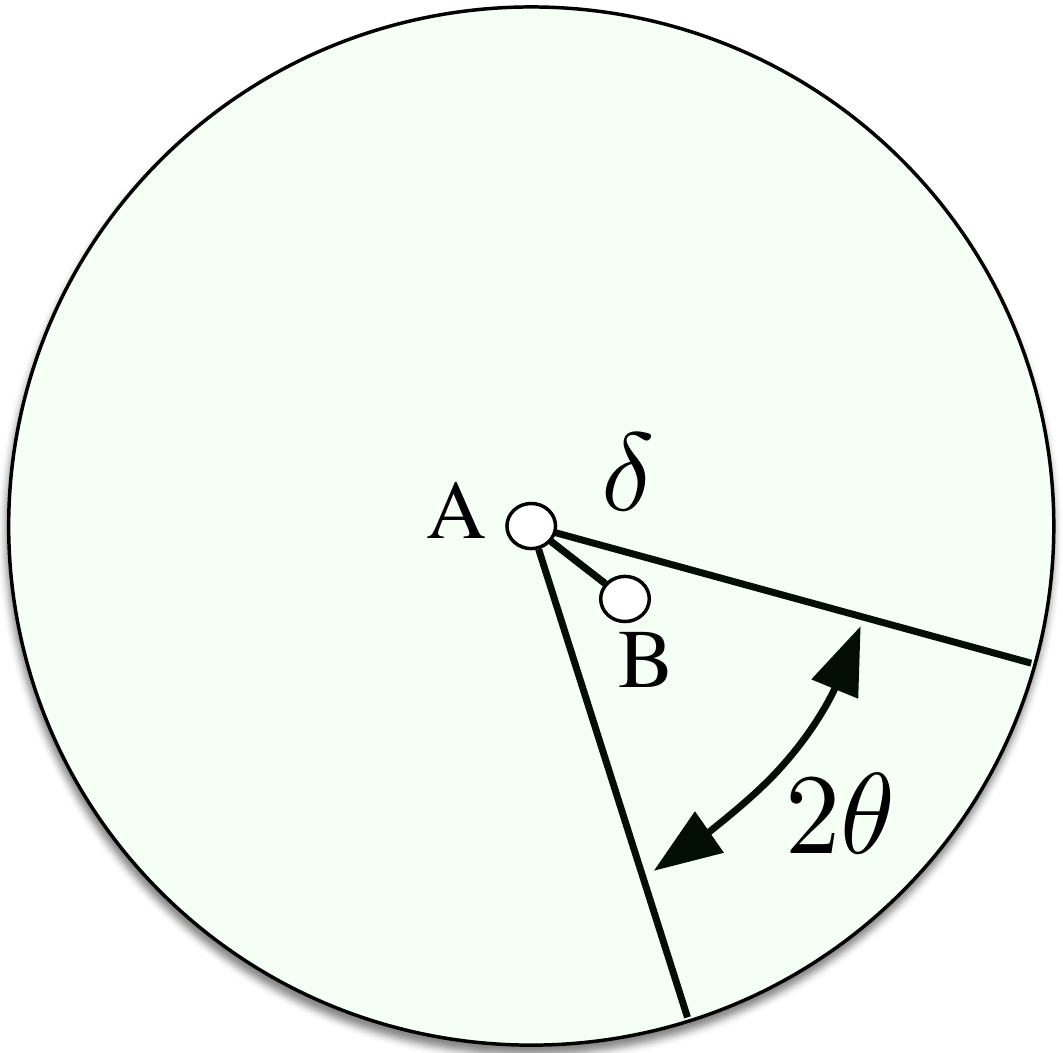}
\caption{ The nodes which are attached to the new hub B occupy a space
  of angle $2\theta$. The average number of nodes in this section is
  $\theta N/\pi$.  }
\label{fig:4}
\end{figure}

Capacity constraints (such as the number of cars on a given road) are
important and result in an increase of time or cost. A popular form
which expresses the travel time $\tau(e)$ on a link of length $d(e)$ is given by the {\it Bureau of Public
  Road} (BPR) function~\cite{BPR:1964} which can be written as
\begin{equation}
\tau(e)=\frac{d(e)}{\overline{v}}\left[1+\left(\frac{T(e)}{q_0}\right)^{\phi}\right]
\label{eq:bpr}
\end{equation}
where $\phi>0$, $\overline{v}$ is the average velocity on links, and
$T(e)$ is the traffic on $e$. In our simplified model, we use the BC
as a simple proxy for the traffic: $T(e)=g(e)$. The quantity $q_0$
corresponds to the practical capacity of the link, and the exponent
$\phi$ is usually large (some empirical studies display a value of
order $\phi\approx 4$ \cite{Branston:1976}) but we will consider here
the lowest non-trivial value $\phi=1$. We consider the problem of
finding the optimal graph that minimizes the total time (which we will
also call `energy') spent on the graph and given by
\begin{align}
\nonumber
{\cal E}(q_0)&=\sum_e g(e)\tau(e)\\
&\propto\sum_eg(e)d(e)\left[1+\frac{g(e)}{q_0}\right]
\label{eq:E}
\end{align}
If the capacity is very large $g(e)\ll q_0$, congestion effects are
absent and we recover an optimal network corresponding to
${\cal E}\approx \sum_eg(e)d(e)$ discussed above ($(\mu,\nu)=(1,1)$)
where centrality dominates over distance and with the presence of a
few dominant hubs. In the opposite case where the capacity is very
small $g(e)\gg q_0$, we obtain an energy of the form
${\cal E}\approx\sum_e g(e)^2d(e)$. Our numerical simulation indicates
that the corresponding optimal network is a `star-hub' where one node
(the hub) is connected directly to all the other nodes. This is also
the situation which corresponds to the shortest path tree obtained by
minimizing the shortest path ${\cal E}\propto\langle\ell\rangle$ (in
the notation introduced above it corresponds to
$(\mu,\nu)=(1,0)$). The main difference here is that the distance term
in Eq.~\eqref{eq:E} will select the hub as being the closest node to
the gravity center of all nodes. When we increase the capacity, the
optimal network will thus evolve from a star configuration to a
spatially organized network. In order to monitor this topological
transition, we can measure different quantities, and we will focus
here on the degree dominance $d^*$ defined as
\begin{equation}
d^*=\frac{1}{(N-1)(N-2)}\sum_i(k^*-k_i)
\end{equation}
where $k^*=\max_i k_i$. For a star network we have $d^*=1$, and for an
homogeneous network with a constant degree, the dominance is
$d^*=0$. The evolution of the degree dominance when $q_0$ is
increasing is shown for numerical simulations in Fig.~\ref{fig:dstar}.
We observe in this figure a sharp transition in the dominance for
$q^*\approx 1$. We can estimate where the transition takes place and
we provide here this simple argument, valid for energies of the form
\begin{equation}
{\cal E}=\sum_eg(e)d(e)\left(1+\frac{g(e)}{q_0}\right)
\end{equation}
Starting from the star graph situation at low $q_0$, the transition occurs when a
new hub appears, as shown in Fig.~\ref{fig:3a}. 
In order to compute $q^*$, we have then to estimate the energy
difference between the one star system and the new two-stars network
where the hubs are separated by some distance $\delta$ (see
Fig.~\ref{fig:3b}).
For the one star system, all centralities are equal and given by
\begin{equation}
g\sim N/N^2\sim 1/N
\end{equation}
(all BCs are normalized here by a factor $1/N^2$). The energy of the one hub (A) configuration is thus
\begin{equation}
{\cal E}\sim \left(\frac{1}{N}+\frac{1}{N^2q_0}\right)\sum_ed(e)
\end{equation}
The length of links correspond to the distance to the hub `A':
$\sum_ed(e)=\sum_{i=1}^N|r_i-r_A|$ where $r_i$ denotes the position of node $i$. Numerical experiments
suggest that a new hub `B' appears (see Figure~\ref{fig:3b}), and we denote the vector between
A and B by $\delta$. The simulations also suggest that the new
hub is connected to a set of $aN$ nodes comprised in an angle
$2\theta$ (see ( Figure~\ref{fig:4}).
For networks large enough, when the density of nodes is uniform, the
angle $\theta$ and the number of nodes can simply be related by $aN=N
2\theta/2\pi$ leading to $a=\theta/\pi$. The links attached directly to A or
B have a centrality which is of order $\sim 1/N$, and only the centrality of
the link $A-B$ is different and is of the order $a(1-a)$. The energy of this new system is then
\begin{eqnarray}
\nonumber
{\cal E}'=\left(\frac{1}{N}+\frac{1}{N^2q_0}\right)
\left[\sum_{i\in A}|r_i-r_A|+\sum_{i\in B}|r_i-r_B|\right]\\
+\delta a(1-a)
\left[1+\frac{a(1-a)}{q_0}\right]
\end{eqnarray}
where $i\in A(B)$ means that the summation is over the nodes directly
attached to A(B). The energy difference $\Delta{\cal E}={\cal
  E}'-{\cal E}$ at lowest
non-trivial order in $\delta$ is then given by
\begin{align}
\nonumber
\Delta{\cal E}&\approx a(1-a)\delta \left(1+\frac{a(1-a)}{q_0}\right)\\
&-\left(\frac{1}{N}+\frac{1}{N^2q_0}\right)\sum_{i\in B}\frac{\delta\cdot
  (r_i-r_A)}{|r_i-r_A|}+{\cal O}(\delta^2)
\end{align}
By using a continuous approximation for computing the sum, we obtain
at dominant order in $\delta$ and $N$
\begin{align}
\nonumber
\Delta{\cal E}\approx
  \delta a(1-a)\left(1+\frac{a(1-a)}{q_0}\right)\\
-\frac{\delta}{\pi}\sin  a\pi
+{\cal O}(\delta^2,\frac{1}{N})
\end{align}
We can easily study this function and show that its sign depends on
the slope at $a=0$. An expansion around this point gives
\begin{align}
\Delta{\cal E}\approx \delta a^2\left(\frac{1}{q_0}-1\right)+{\cal O}(a^3)
\end{align}
which shows that there is a transition for $q_0=q^*=1$. If $q_0>1$,
$\Delta{\cal E}<0$ and a second hub can appear. This simple argument thus predicts that
for large enough networks, the one star configuration is indeed stable
up to a finite value of the capacity $q_0$, and also predicts a value $1.0$ which is in good agreement with the
value observed in numerical simulations.

We thus see in this simple toy model that congestion can induce
transitions between the structure of networks. For large capacity,
congestion is irrelevant while for smaller capacity, we observe a
transition to a centralized organization with one main hub. Also, this
formalism discussed here probably allows for other, more general
studies about the effect of congestion in optimal networks. In
particular, we could probably study the effect of a more general cost
function (Eq.~\eqref{eq:bpr}), or at least the effect of other values of the
exponent $\phi$.


\subsection{Fluctuations and the emergence of loops}



In most examples studied in the literature, optimal networks are trees. However in many natural
networks such as veins in leaves or insect wings, one observes many
loops. In \cite{Banavar:2000}, the authors show that loops can emerge
in optimal networks according to the convexity of the cost
function. More precisely they consider the total transportation cost
in the network given by
\begin{align}
{\cal E}=\sum_ek_e|i_e|^\gamma
\end{align}
where $i_e$ is the quantity of material along the link $e$, $k_e$ is
the resistance and $\gamma$ is an exponent (in the electric circuit
case $\gamma=2$, $i_e$ is the current, and ${\cal E}$ is the power
dissipated). In addition, the total flow of injected currents is
assumed to be equal to the outflow. For $\gamma>1$, there is then a unique
flow pattern with non-zero current along all links (and therefore with
loops). In contrast, for $0<\gamma<1$, the solution is not unique but
all of them have a tree structure \cite{Banavar:2000}. These results
were confirmed in \cite{Bohn:2007} where the authors show a transition
from trees to graphs with loops and which corresponds to a
discontinuity in the slope of the cost function.

More recently, two studies which appeared simultaneously
\cite{Corson:2010,Katifori:2010} rediscussed the problem of the
existence of a non-zero (and sometimes high) density of loops in real
optimal networks such as veination patterns in leaves. In particular,
it seems that for real-world systems the existence of fluctuations is
crucial in the formation of loops. In the context of the evolution of
leaves, the resilience to damage also naturally induces a high density
of loops (see \cite{Katifori:2010} for an example of flow re-routing
after an injury).

In these studies, the model is defined on an electrical network with
conductances $C_{e}$ on each link and the total dissipated power 
\begin{align}
P=\frac{1}{2}\sum_k\sum_{j\in\Gamma(k)}C_{kj}(V_k-V_j)^2
\end{align}
is minimized under the cost condition
\begin{align}
\frac{1}{2}\sum_k\sum_{j\in\Gamma(k)}C_{jk}^{\gamma}=1
\end{align}
where in this equation it is assumed that the cost of a conductance
$C_{kj}$ is given by $C_{kj}^{\gamma}$ where $\gamma$ is a real
number. This constraint can be interpreted as the limit of 
the amount of resources available to construct the network. 
The quantity $V_i$ is the potential at node $i$.

Following \cite{Katifori:2010}, we can introduce two variants of this
model. The first one which represents the resilience to damage is
defined as follows. We cut a link $e$ and compute the
total dissipated power denoted by $P^{e}$. The resilience to this
damage can then be rephrased as the minimization of the 
functional
\begin{align}
R=\sum_{e\in E} P^e
\end{align}
Note that if breaking the link $e$ disconnects the network, it will
lead to an infinite resistance and to an infinite value of the dissipated
power $P^e$: the finiteness of R implies the existence of
loops in the optimal network. In another variant, \cite{Katifori:2010}
Katifori et al. introduces time fluctuating load by considering a system with
one source at the stem of the leaf and one single moving sink at
position $a$. For this system, one can compute the total dissipated
power $P^a$ and the resilience to fluctuations can be rephrased as the
minimization of the functional
\begin{align}
F=\sum_aP^a
\end{align}

\begin{figure}[h!]
\includegraphics[width=0.50\textwidth]{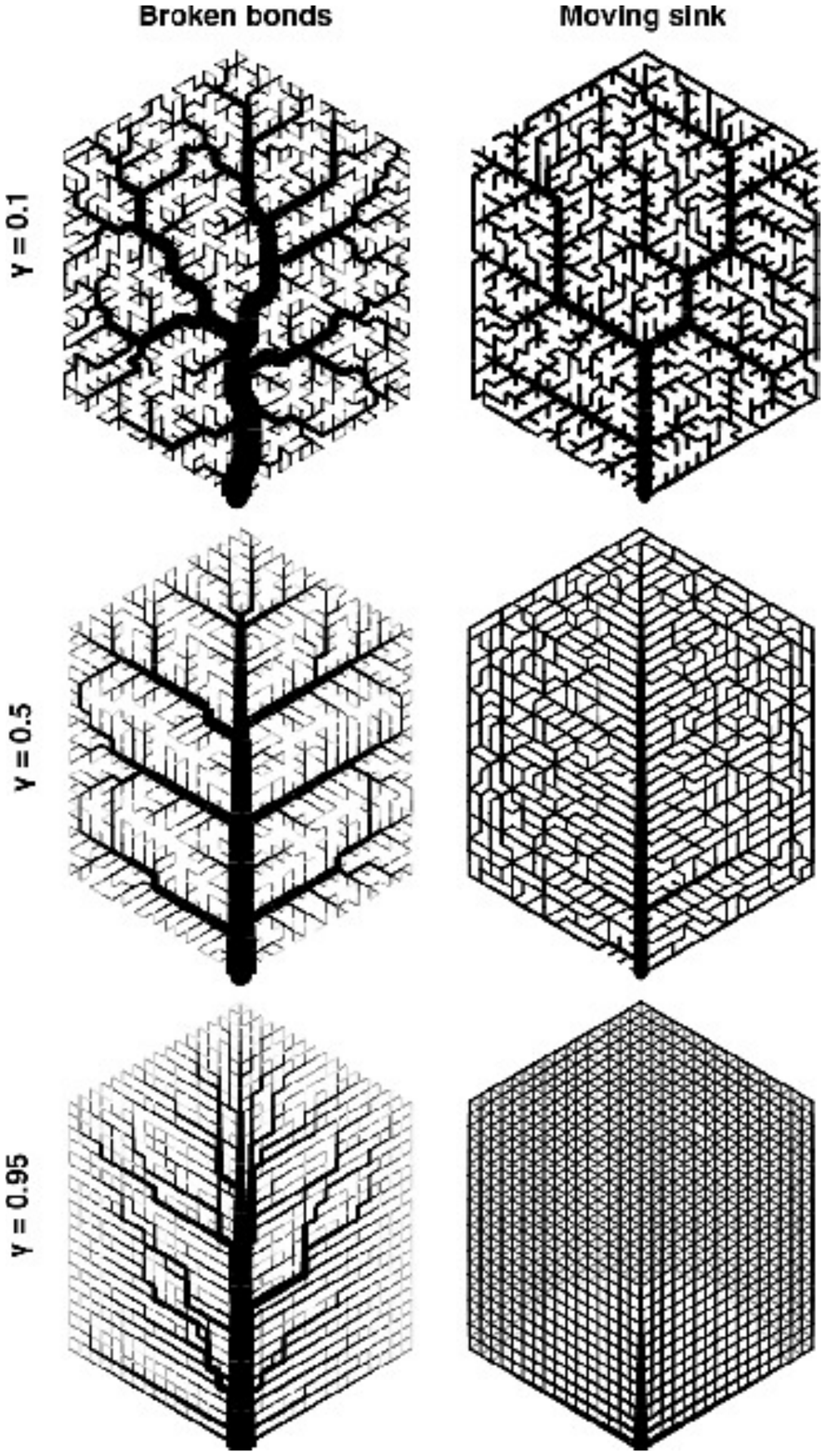}
\caption{ Optimal networks obtained for different values of $\gamma$
  and for (left panels) the resilience to damage and (right panels)
  for a fluctuating load. We observe loops ($\gamma<1$) and different
  structures according to $\gamma$. The thickness of links is a
  function of their conductance. Figure taken from
  \cite{Katifori:2010}. }
\label{fig:katifori}
\end{figure}

In these models, one observes the formation of loops (see
Fig.~\ref{fig:katifori}), reminiscent of the ones seen in real
leaves. These studies shed a new light in the formation and the
evolution of real-world networks and open interesting directions of
research. In particular, it would be very interesting to understand
more quantitatively the condition of appearance of loops in spatial networks.

\section{Discussion}



We tried here to review some examples of transitions in spatial
networks. We didn't mean to be exhaustive and other models that
display transitions or crossovers could have been discussed here. We
saw that in some cases the transition can be percolation-like with the
appearance of a giant cluster. In other cases we observe less common
transitions with different phases that are characterized by a
different scaling of some quantity with the system size (the average
shortest path or the average link length, etc.). These `topological
transitions' between different structures can be of interest in
real-world situations where it is not only the total connectivity that
can vary (as it is the case for percolation) but other features
connected to navigation over these networks. Some transitions can also
be more subtle and concerns essentially the traffic on networks and
the results about the localization of bottlenecks shed light on the
organizaton of flows in spatial networks.  In the different models that we
described here, the spatial distribution of the betweenness centrality
can indeed display some sort of localization transition when the density of edges
increases with a concentration of bottlenecks around the gravity
center of the system. Generally speaking, the study of high BC nodes is an important
endeavor as it represents a generalization of studying the maximum BC
nodes that governs the behavior of the system in saturated cases
where the traffic exceeds the capacity of links.

All these examples show that spatial networks display a large variety of behaviors with
crossovers or transitions that separate different regimes
characterized by different large-scale properties. A more systematic
study would be needed here in order to distinguish sharp transitions
from crossovers and to understand if these phenomena can be understood
in a more general framework, such as phase transitions in statistical
physics.

\medskip
{\bf Acknowledgments}

  I thank my colleagues and collaborators for many interesting
  discussions on various aspects of spatial networks: H. Barbosa, M. Batty, H. Berestycki,
  P. Bordin, J. Bouttier, C.P. Dettmann, P. Di Francesco, A. Flammini, G. Ghoshal,
  E. Guitter, P. Jensen, A. Kartun-Giles, E. Katifori, A. Kirkley,  V. Latora, T. Louail, J.-M. Luck,
  J.-P. Nadal, V. Nicosia, K. Mallick, C. Roth, E. Strano,
  M.P. Viana. I also thank the geohistorical data group, in particular
  M. Gribaudi and J. Perret. I thank P. Grassberger for pointing out
  some mistakes and important references.



\bibliographystyle{apsrev}
\bibliography{bibfile_physrep}		         

\end{document}